\documentclass[12pt,preprint]{aastex} 
\newcommand{\mrk}{Mrk~279} 
\newcommand{\kms}{km~s$^{-1}$} 
\newcommand{\lam}{$\lambda$} 
\newcommand{\lya}{Ly$\alpha$} 
\newcommand{\fuse}{{\it FUSE}} 
\newcommand{\stis}{STIS} 
\newcommand{\chand}{{\it Chandra}}
\newcommand{\xmm}{{\it XMM-Newton}}

\begin{document}

\title{Variable Intrinsic Absorption in Mrk~279}

\author{Jennifer E.\ Scott\altaffilmark{1,2}, 
Nahum Arav\altaffilmark{3,4},
Jack R.\ Gabel\altaffilmark{3,5},
Gerard A.\ Kriss\altaffilmark{6,7},
Jessica Kim Quijano\altaffilmark{6},
Jelle S.\ Kaastra\altaffilmark{8},
Elisa Costantini\altaffilmark{8}
Kirk Korista\altaffilmark{9}}

\altaffiltext{1}{
Observational Cosmology Laboratory,
National Aeronautics and Space Administration,
Goddard Space Flight Center,
Greenbelt, MD 20771  USA}

\altaffiltext{2}{
present address:  Department of Physics, Astronomy, and Geosciences, Towson University,
Towson, MD  21252 USA; jescott@towson.edu}

\altaffiltext{3}{Center for Astrophysics and Space Astronomy, University
of Colorado, 389 UCB, Boulder, CO  80309  USA}

\altaffiltext{4}{
present address: Department of Physics, Virginia Polytechnic Institute \& State University,
Blacksburg, VA  24061 USA;
arav@vt.edu}

\altaffiltext{5}{
present address: Department of Physics,
Creighton University,
Omaha, NE  68178 USA;
jackgabel@creighton.edu}

\altaffiltext{6}{Space Telescope Science Institute, 3700 San Martin Drive,
Baltimore, MD  21218 USA; [gak,jkim]@stsci.edu}

\altaffiltext{7}{Center for Astrophysical Sciences, Department of 
Physics and Astronomy,
The Johns Hopkins University, Baltimore, MD 21218 USA} 

\altaffiltext{8}{SRON National Institute for Space Research,
Sorbonnelaan 2, 3584 CA Utrecht, The Netherlands; [J.S.Kaastra,
e.costantini]@sron.nl}

\altaffiltext{9}{Department of Physics, Western Michigan University, Kalamazoo, MI 49008 USA; 
korista@wmich.edu}

\addtocounter{footnote}{-7}

\begin{abstract}
We examine the variability in the
intrinsic absorption in the Seyfert 1 galaxy \mrk\
using three epochs of observations
from the {\it Far Ultraviolet Spectroscopic Explorer (FUSE)} and two
epochs of observations with the Space Telescope Imaging Spectrograph
on the {\it Hubble Space Telescope}.   
Rather than finding simple photoionization responses of the absorbing gas to changes in the
underlying continuum, the observed changes in the absorption profiles 
can be understood more clearly
if the effective covering fraction of the gas in all emission components,
continuum and broad and intermediate velocity width emission lines, is
accounted for.
While we do not uniquely solve for all of these separate covering fractions
and the ionic column densities using the spectral data, we examine the parameter space 
using previously well-constrained solutions for continuum and single emission 
component covering fractions.  Assuming full coverage of the continuum, we find
that of the two velocity components of the Mrk~279 absorption most likely associated with
its outflow,
one likely has zero coverage of the intermediate line region while the
other does not.  For each component, however, the broad line region is more
fully covered than the intermediate line region.
Changes in the \ion{O}{6} column densities are unconstrained due to
saturation, but
we show that small changes in the nonsaturated \ion{C}{4} and \ion{N}{5}
column densities 
are consistent with the outflow gas having zero or partial
covering of the intermediate line region and an ionization 
parameter changing from $\sim$0.01 to $\sim$0.1 from 2002 to 2003
as the UV continuum flux increased by a factor
of $\sim$8.
The  absence of a change in the \ion{C}{3}
absorbing column density is attributed to this species arising outside 
the Mrk279 outflow.
\end{abstract}


\section{Introduction}

Mass outflows from low-redshift active galactic nuclei (AGNs)
are inferred from blueshifted, variable X-ray and UV absorption
which often shows evidence for partial coverage of the contintuum source
in nearly one half of all nearby Seyfert galaxies   
(see review in Crenshaw, Kraemer, \& George 2003 and recent results in Dunn et al.\ 2007).
In addition to possibly playing a strong role in the 
accretion process 
(Blandford \& Begelman 1999, 2004),
AGN outflows likely have substantial impacts on 
their environments, providing feedback energy
to regulate the formation of galaxies and clusters of 
galaxies
(Granato et al.\ 2004, Scannapieco \& Oh 2004, Scannapieco, Silk, \& Bouwens 2007);
to set the luminosity-temperature relation in galaxy
clusters (Cavaliere et al.\ 2002); 
to distribute metals in galaxy clusters (Moll et al.\ 2007) and into the intergalactic
medium (Adelberger et al.\ 2003).  

Theoretical models of AGN outflows, from accretion disk winds
(K\"{o}nigl \& Kartje 1994, Murray et al.\ 1995, Proga 2000, 2003,
Proga, Stone, \& Kallman 2000, Proga \& Kallman 2004),
or from  ablation from the obscuring torus
(Krolik \& Kriss 1995, 2001)
place the absorbing material at different distances from  
the central engine and with different orientations with respect to the
broad line region (BLR).
A large sample of high resolution X-ray and UV spectra 
of ouflows in nearby Seyferts from
{\it Chandra}, \xmm,
the {\it Far Ultraviolet Spectroscopic Explorer (FUSE)}
and from 
the Space Telescope Imaging Spectrograph (STIS) onboard
the {\it Hubble Space Telescope}
has been building 
in the literature over the last decade,
providing much insight into the physical conditions in
these ouflows. 
However, a broad understanding of these outflows has yet to emerge, and 
several outstanding questions remain, namely:
What is the relationship between
the X-ray- and UV- absorbing gas?;
What is the location and structure of the absorbing material?;
Is the observed variability in some absorption profiles 
a response to flux variations in
the central engine 
(Krolik \& Kriss 1995, 
Shields \& Hamann 1997, Crenshaw et al.\ 2000, Kraemer et
al.\ 2002)
or to bulk motions of the absorbing gas (Crenshaw \& Kraemer 1999, 
Kraemer et al.\ 2001, Gabel et al.\ 2003b)?

In general, it is difficult to associate a particular UV absorption component
with the X-ray absorber unambiguously, in part due to relatively low spectral resolution 
in X-ray data and in part because the ionization modeling
is sensitive to the shape of the UV-to-X-ray continuum (Kaspi et al.\ 2001).
However, with  the
exception of campaigns on Mrk~279, using \fuse, \stis, and \chand\ with
both the {\it High Energy Transmission Grating (HETG)} 
(Scott et al.\ 2004, Gabel et al.\ 2005a, S04 and G05 hereafter) and the
{\it Low Energy Transmission Grating} 
(Costantini et al.\ 2007), on NGC~7469 (Scott et al.\ 2005)
using \fuse, \stis, and \chand/{\it HETG},
the long-term observing campaign on NGC~3783, using
\fuse, \stis, and \chand/{\it HETG}
(Kaspi et al.\ 2002, Netzer et al.\ 2003, Gabel et al.\ 2003a, 2005b),
and the campaign on NGC~4151 using \fuse, \stis, and \chand/{\it HETG}
(Kraemer et al.\ 2005, Kraemer et al. 2006, Crenshaw \& Kraemer 2007),
UV and X-ray observations are generally obtained at different times.

Several observations of the absorption profiles  taken over 
a period of time
are particularly useful for addressing the last two questions
mentioned above,
as they provide contraints on the response time of the gas
to variability in the central engine and thus limits on the
distance of the absorber from the continuum-emitting region.
Mrk~279 ($z_{em}=0.0305$) has been the subject of an intensive
campaign of simultaneous observations in the UV and X-ray
(S04, Kaastra et al.\ 2004, Arav et al.\ 2005, 2007,
G05, Costantini et al.\ 2007).
In this paper, we use several epochs of UV spectral obsevations
of Mrk~279 over 1999 to 2003 from
\fuse\ and \stis\ presented in  S04 and G05
to investigate the variability of the 
column density and/or covering fraction of the outflow over time.
We choose to concentrate on the UV data because the 
2002 epoch of \chand\ observations coincided with a low
flux state.  The grating data therefore  have signal-to-noise
(S/N) too low to perform detailed analyses (S04).

The spectral coverage of
\fuse\ provides information on \ion{H}{1} absorption in Ly~$\beta$
and higher order
Lyman transitions, and \ion{C}{3}~\lam977,
while 
the \stis\ observations cover Ly$\alpha$ and the \ion{N}{5},
\ion{Si}{4}, and \ion{C}{4} doublets.
The different epochs permit us to investigate the
effects of UV continuum variability on the intrinsic absorption.  
The analysis presented
here will also take into account changes in absorption profiles that
may be a result of differing contributions of the various emission
sources, i.e. continuum versus broad lines, in different observation epochs.

\section{Continuum and Emission Lines}
\label{sec-cont}

\begin{figure}
\epsscale{0.8}
\plotone{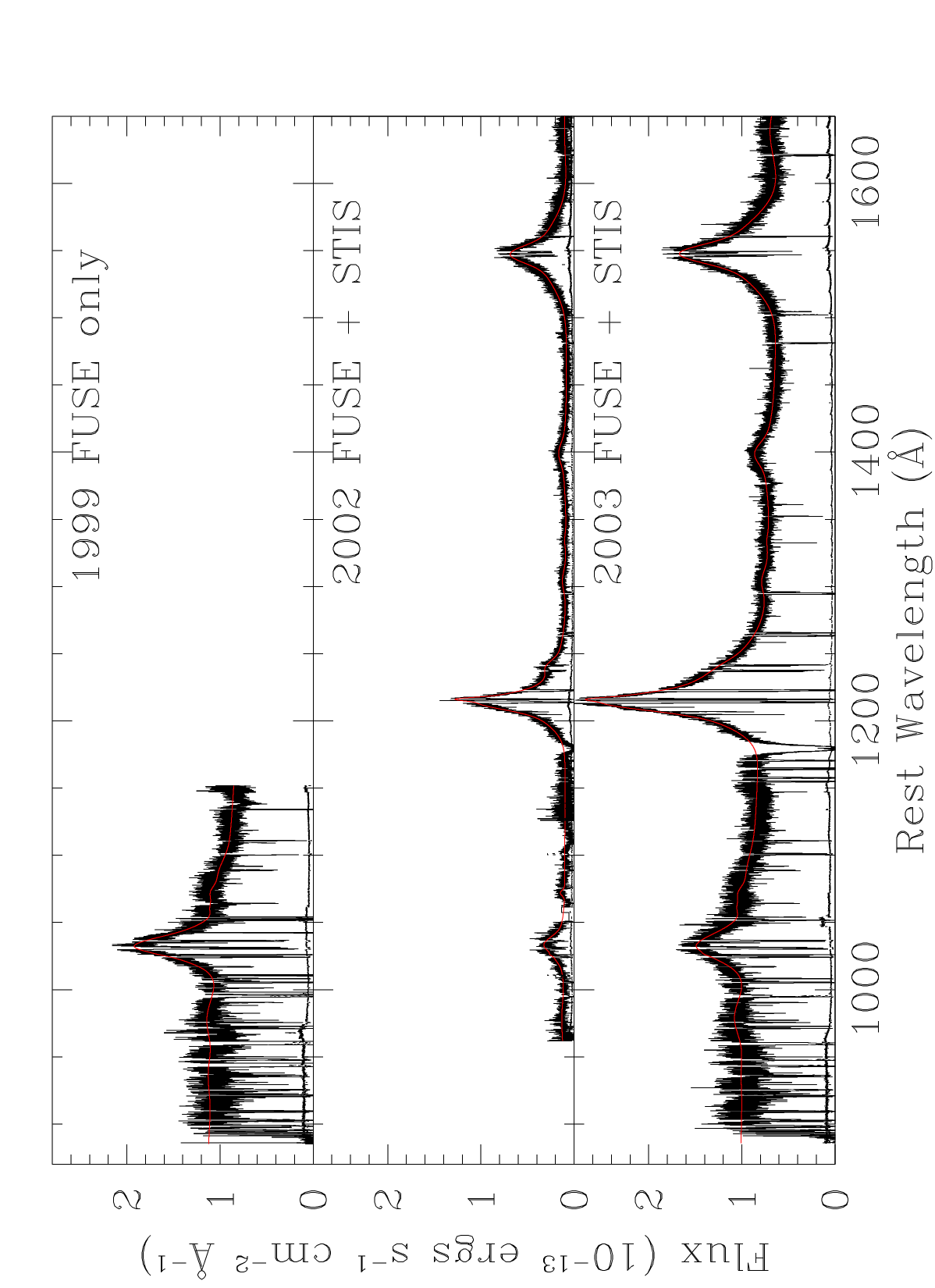}
\caption{1999 \fuse\ spectrum (top) and 2002 and 2003
\fuse\ + \stis\ spectra (middle and bottom) of \mrk\ with
errors (dotted line) and continuum and emission line fits (red)
\label{fig:specall}}
\end{figure}

Table~\ref{table-data} summarizes the data used in this analysis. 
See S04 and G05 for full presentations of
the \fuse\ and \stis\ spectra.
The 1999 \fuse\ spectrum and the 2002 and 2003 \fuse\ + \stis\
spectra are shown in Figure~\ref{fig:specall}.

\begin{figure}
\epsscale{0.9}
\plotone{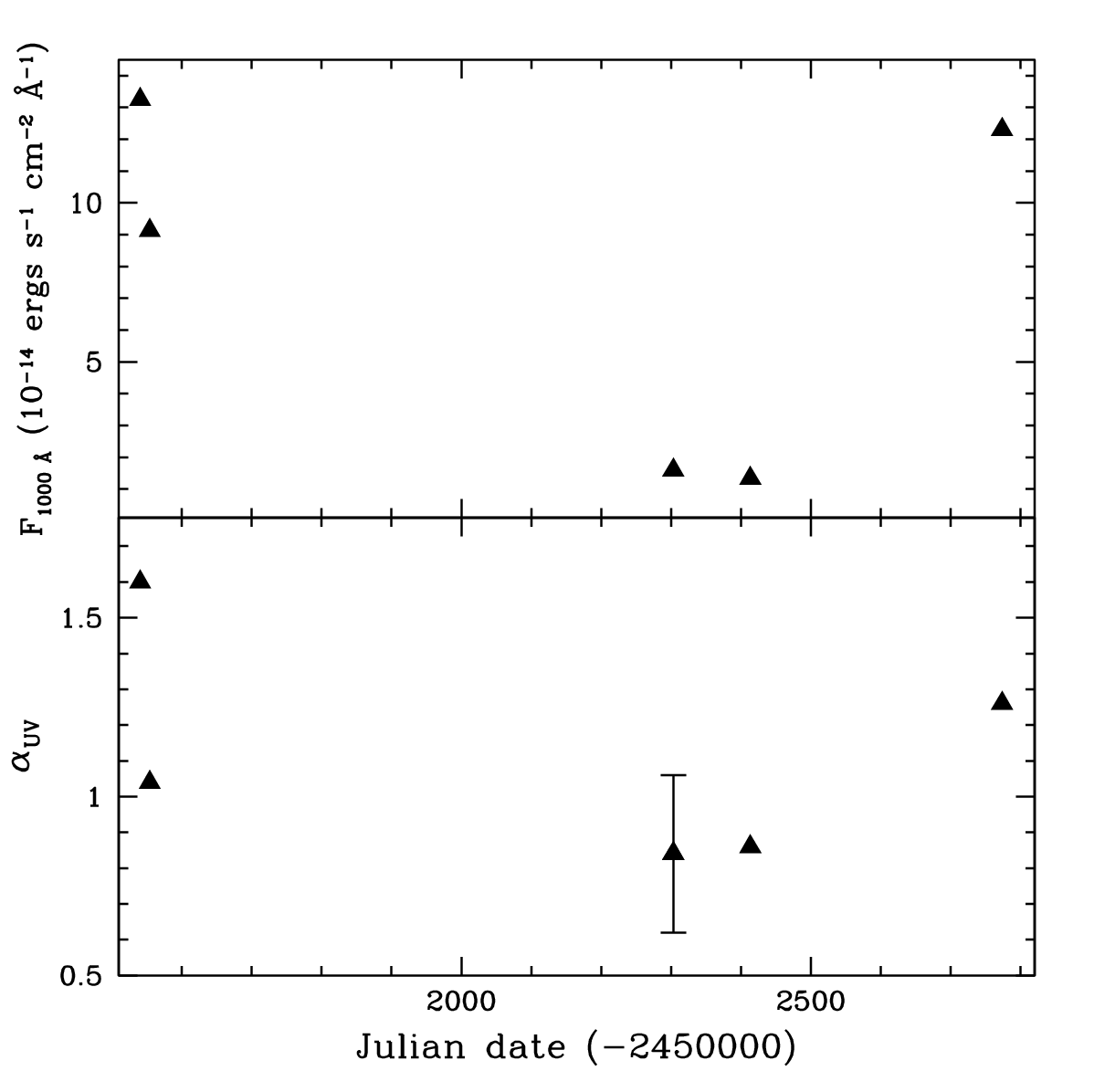}
\caption{Variation in continuum flux and UV spectral index 1999-2003
\label{fig:varflux}}
\end{figure}

We fit a power law continuum of the form
$f_{\lambda} = f_{1000} \left(\frac{\lambda}{1000 \; {\rm \AA}}\right)^{-\alpha}$ to
the 2003 \fuse\ and \stis\ spectra simultaneously, where
the normalization, $f_{1000}$, is the flux at 1000 \AA, and we
include
the Galactic extinction
law of Cardelli, Clayton \& Mathis (1989) with R$_{V}=3.1$ and
E~(~B~-~V~)~=~0.016 (Schlegel, Finkbeiner, \& Davis 1998) in the continuum fits.
The best fit continuum slope and normalization are $\alpha=1.261\pm0.005$ and
$f_{1000}=1.230\pm0.002 \times 10^{-13}$ ergs s$^{-1}$ cm$^{-2}$ \AA$^{-1}$.
For the 1999- and 2002-epoch data we found $(\alpha,f_{1000})=
(1.60\pm0.02,1.325\pm0.004\times10^{-13})$ and
$(0.86\pm0.02,0.133\pm0.001\times10^{-13})$, respectively.
These results are shown graphically in Figure~\ref{fig:varflux}.
The 1999 and 2003 epochs are high flux states, and the 
continuum becomes bluer as it brightens.

We fit Gaussian profiles to the emission lines in the combined \fuse+\stis\
spectrum.
The emission lines we identify and fit are listed in Table~\ref{table-emspec}.
See Table~3 of S04 for the fits  to the 1999- and 2002-epoch spectra.
An F-test confirms that the most prominent features require contributions from both a broad velocity
component (BLR, FWHM$\sim7500-10000$ \kms) and an intermediate one (ILR, FWHM$\sim3000$ \kms).
The resulting values of the reduced chi-squared, $\chi^2_\nu=1.6$ and $1.5$, for 2002 and
2003 data, respectively, are significantly improved from the values of $\sim$2 achieved
with no ILR component.  We do not attempt to add any more components to the fit to
improve the $\chi^2_\nu$ because there are no remaining systematics in the fit
residuals and because 
these added components would not be well-motivated physically.
The full spectra with fits are shown in Figure~\ref{fig:specall}.

We note that some of the fitted emission lines show velocity shifts of hundreds of
\kms\ from their 2002 positions.   We performed an emission line fit for the 2003
data with the line centroids tied to their 2002 values.  This led to a larger 
$\chi^2_\nu$ in the final fit (1.6 versus 1.5), but did not change the results of
the absorption line analysis presented below.  Since we will treat our
emission model as phenomenological and because S04 did not constrain the
line centroids in this way (e.g. in fitting the 1999, 2000, and 2002 epoch spectra),
we will use the emission model with the emission line velocity shifts in place.

\begin{figure}
\epsscale{0.8}
\plotone{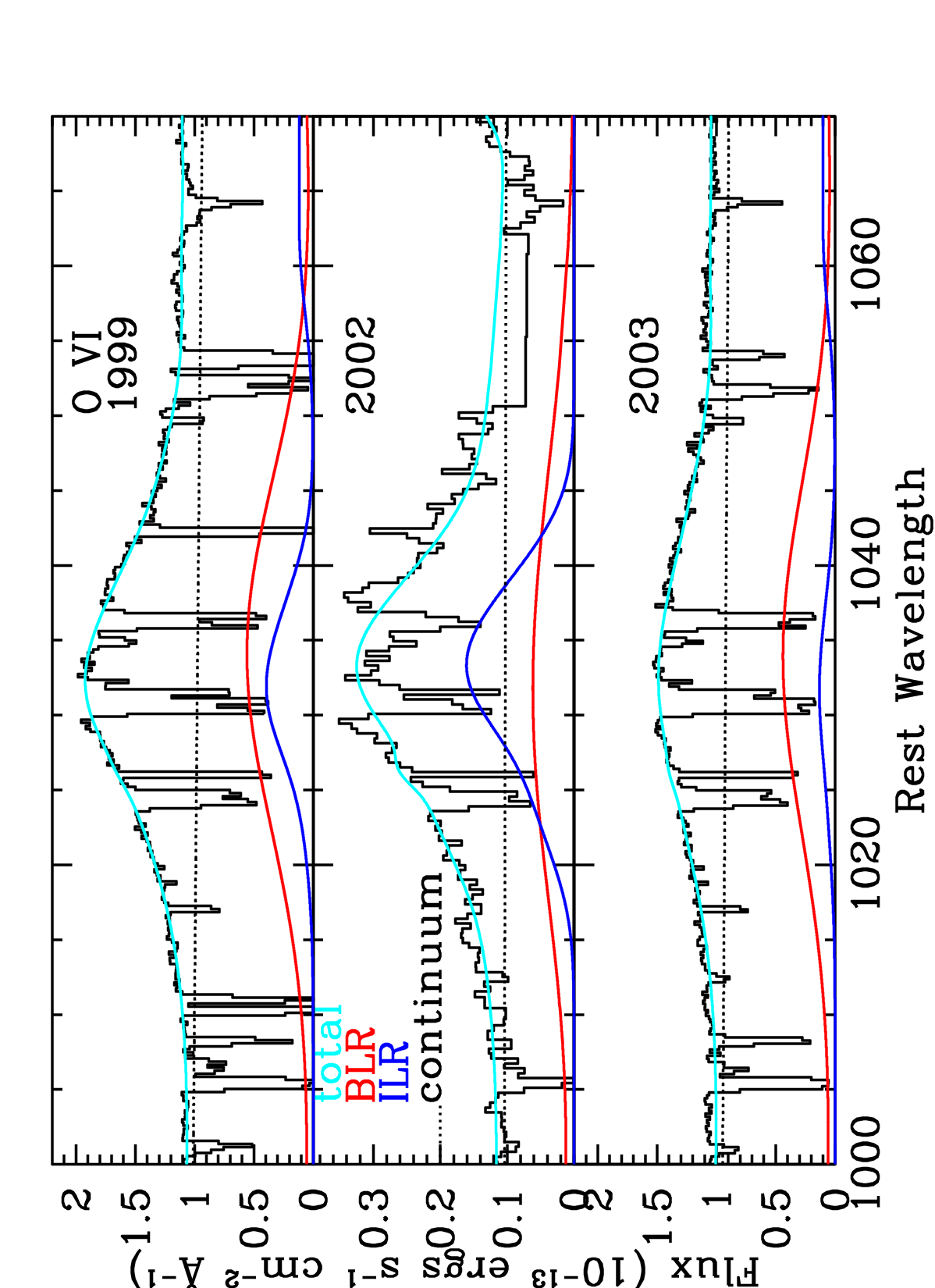}
\caption{\ion{O}{6} emission line profile in 1999, 2002, and 2003 \fuse\ spectra (cyan) with
individual broad line (red) and intermediate line (blue) components and continuum (dotted line) 
shown. Note the expanded scale used to display the 2002 spectrum (middle panel). 
All fitted emission lines are listed in Table~\ref{table-emspec}.
\label{fig:o6_em}}
\end{figure}

\begin{figure}
\epsscale{0.8}
\plotone{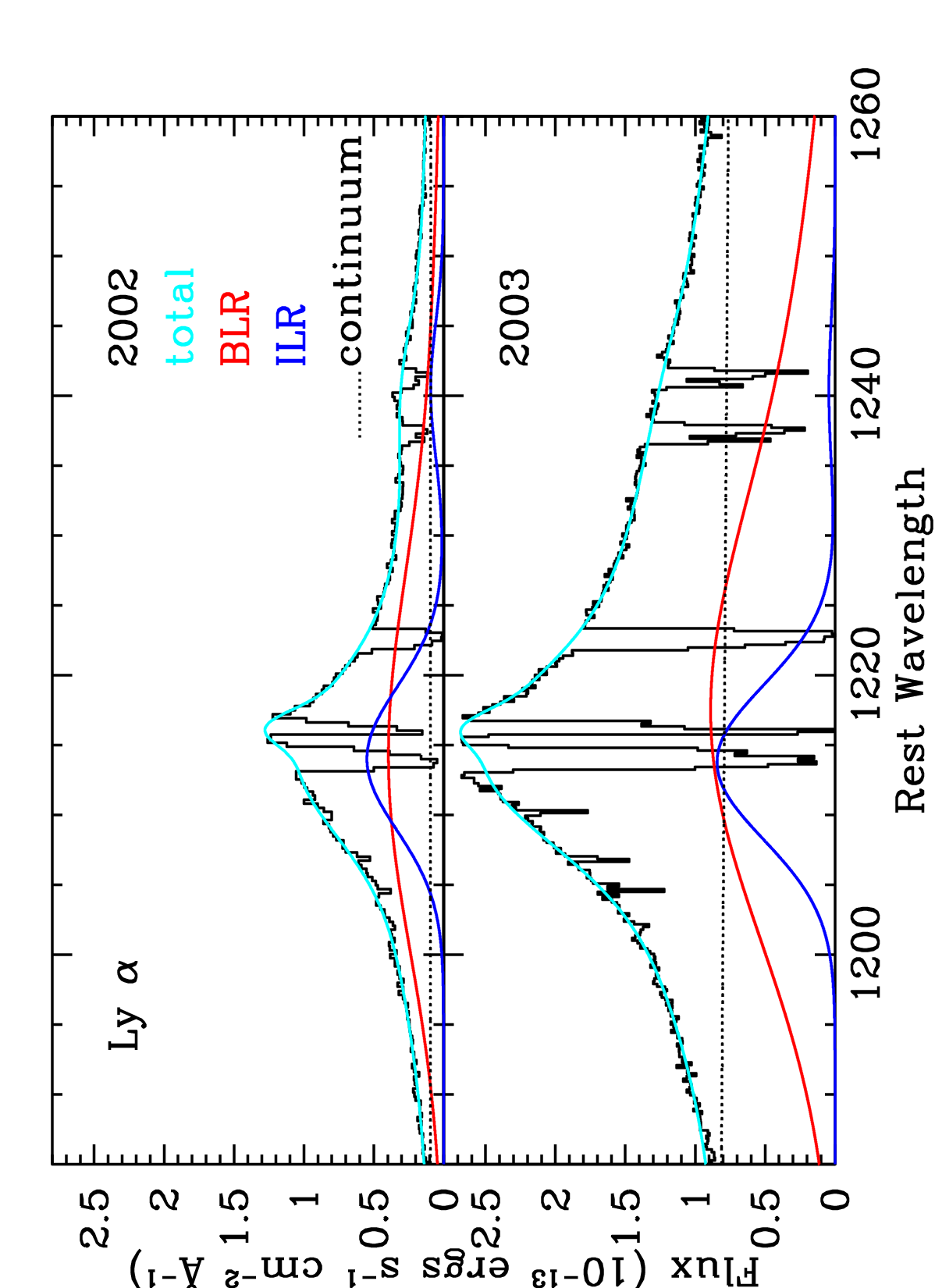}
\caption{Ly$\alpha$ emission line profile in 2002 and 2003 \stis\ spectra (cyan) with
individual broad line (red) and intermediate line (blue) components and continuum (dotted line) 
shown.  The narrow velocity component of the fit (FWHM=697 km s$^{-1}$, fixed to the 2002 value) 
is not shown separately
here.
\label{fig:lya_em}}
\end{figure}

\begin{figure}
\epsscale{0.8}
\plotone{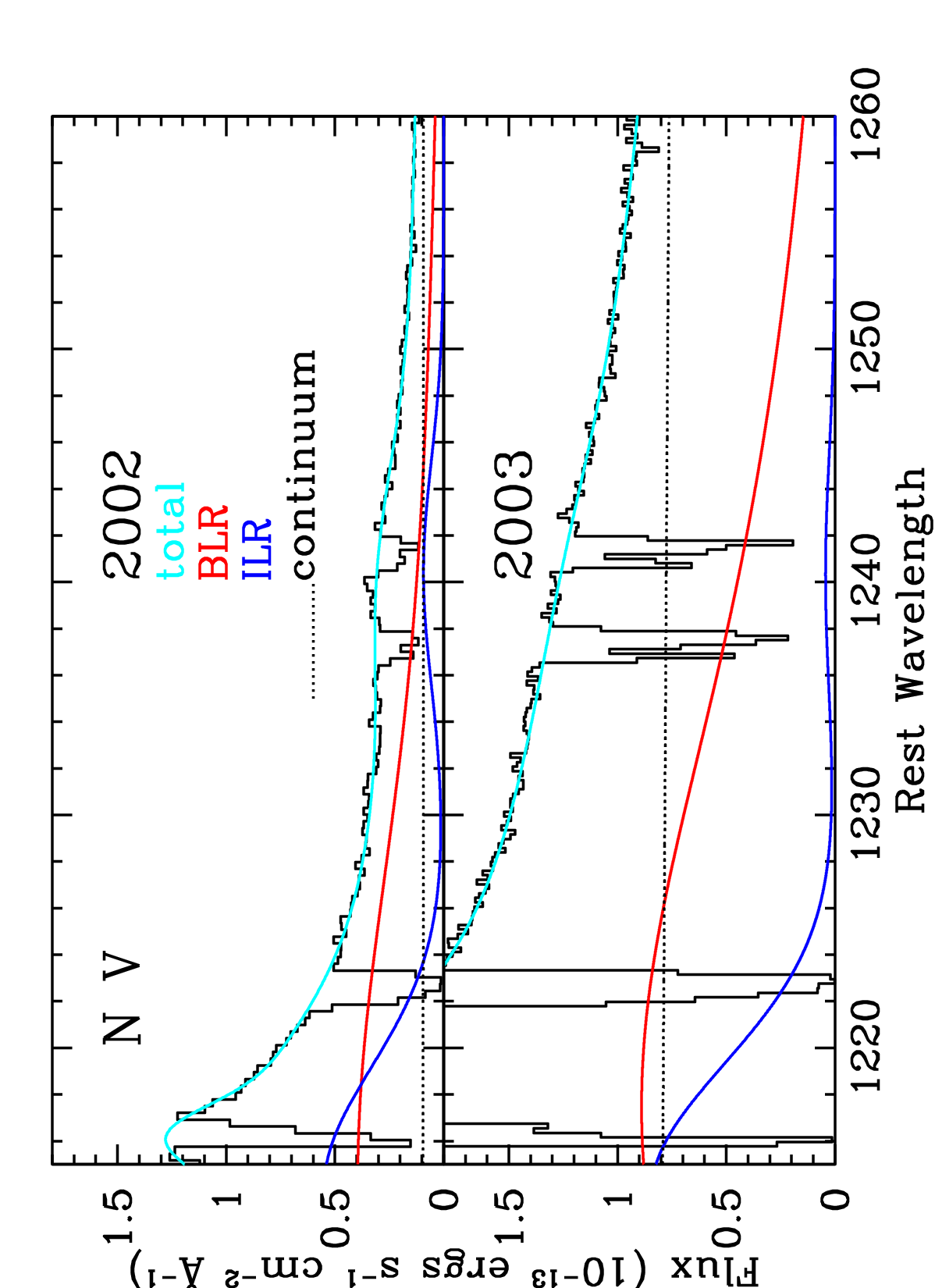}
\caption{\ion{N}{5} emission line profile in 2002 and 2003 \stis\ spectra (cyan) with
individual broad line (red) and intermediate line (blue) components and continuum (dotted line) shown.
\label{fig:n5_em}}
\end{figure}

\begin{figure}
\epsscale{0.8}
\plotone{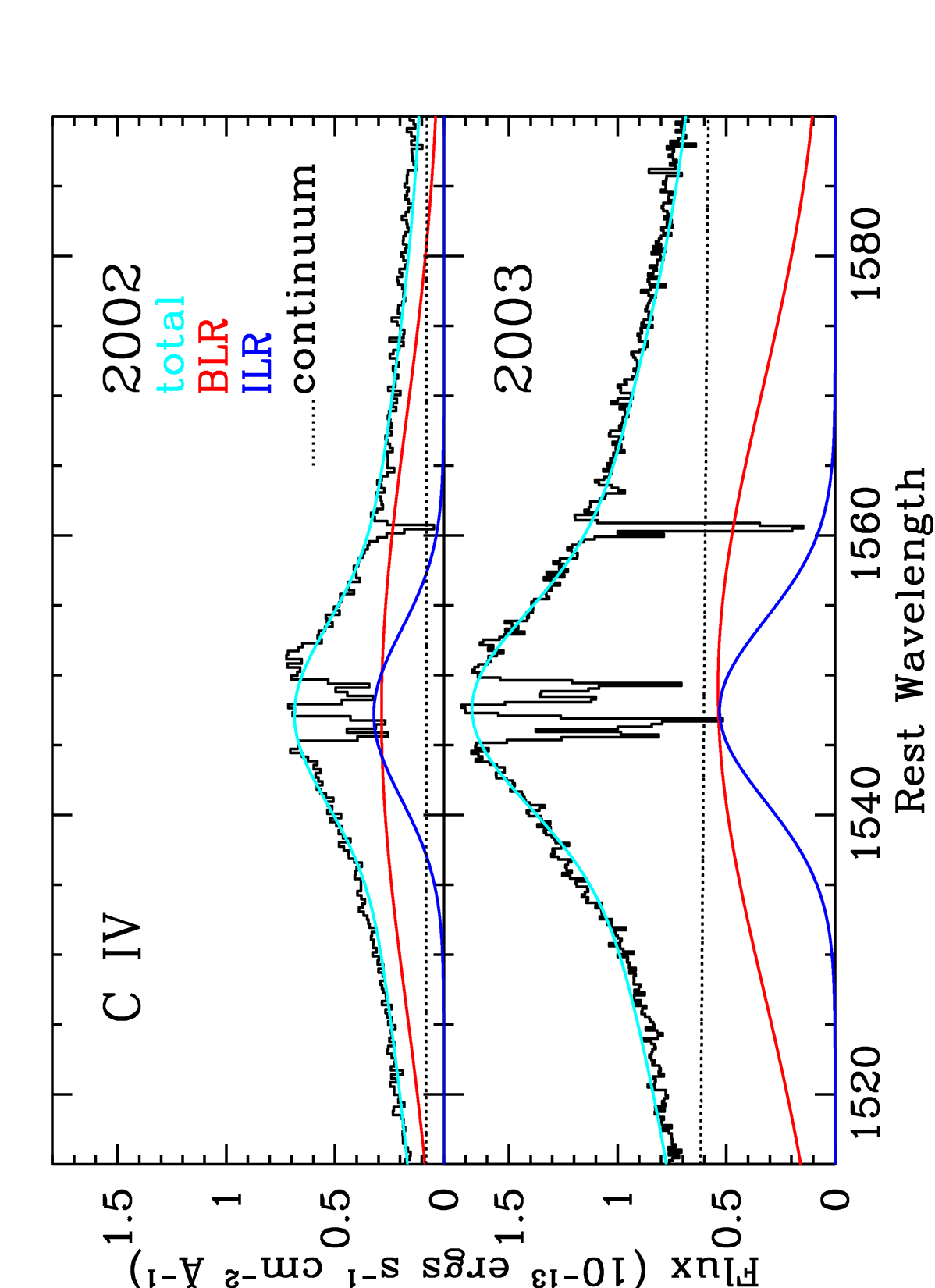}
\caption{\ion{C}{4} emission line profile in 2002 and 2003 \stis\ spectra (cyan) with
individual broad line (red) and intermediate line (blue) components and continuum (dotted line) shown.
\label{fig:c4_em}}
\end{figure}

We show the emission profiles for the 2002 and 2003 spectra in 
Figures~\ref{fig:o6_em}-\ref{fig:c4_em}.
For all these ions, the continuum flux is clearly stronger relative
to the emission lines in the high flux states in 1999 and 2003.
The \ion{O}{6} intermediate line emission is stronger relative to the 
broad line in 2002 than in 1999 or in 2003, when the continuum fluxes
were higher. In fact, the ILR component is very weak in the 2003 spectrum.
The same is true for the Ly$\alpha$ and \ion{N}{5} intermediate lines.  They too are
stronger relative to their 
respective broad lines in the 2002 low flux state than in 2003, the only two
epochs we can compare with \stis\ data.   For \ion{C}{4}, the difference between the 
low and high flux states is less pronounced.
Figure~\ref{fig:c4_em} shows that 
the intermediate width line is only slightly more prominent relative
to the broad line in 2002 than in 2003.

\section{Intrinsic Absorption}

S04 identified several velocity components to the absorption in the \mrk\ spectrum
extending over -600 km s$^{-1}$ - +100 km s$^{-1}$  with respect to the systemic
redshift of the AGN, taken to be $z=0.0305\pm0.0003$.  These components, based on
the velocity structure in the Ly$\beta$ profile, were labeled 1, 2, 2a, 2b, 3, 4, 4a, and 5.
G05 added a component 2c to this,
based on additional velocity structure in the \ion{C}{4} and \ion{N}{5} absorption.
S04 concluded that Components 1 and 4 ($v \sim +90$ km s$^{-1}$ and $-450$ km s$^{-1}$
in Ly$\beta$, respectively)
are likely associated with the disk and possibly the halo of the host galaxy of the AGN and/or
its companion, MCG +12-13-024,
while the origins of Components 3 and 5 ($v \sim -385$ km s$^{-1}$ and $-540$ km s$^{-1}$ 
in Ly$\beta$, respectively)
are ambiguous. The absorption from 
highly ionized species over the entire velocity range
is taken to be intrinsic by G05, as is the \ion{H}{1} absorption that is uncontaminated by any
possible contribution from low-ionization components seen in \ion{C}{2}, \ion{C}{3}, \ion{N}{3},
\ion{Si}{2}, and \ion{Si}{3}, i.e. $v \approx 200-300$ km s$^{-1}$.
We will follow that convention in this paper, referring to Components 2 and 4 in the
highly ionized species to be the absorption complexes centered at $v \approx -300$
and $v \approx -480$ respectively.

The most general form of the intrinsic absorption profile for a particular species 
accounts not only for partial covering, but also for multiple emission sources:
\begin{equation}
\label{equ:int}
I_j = \sum_i R^i_j (C^i_j e^{-\tau_j} + 1 - C^i_j)
\end{equation}
where $I$ is the line intensity at velocity bin $j$, $R$ is the ratio of
source $i$ to the total flux at $j$,
$\tau$ is the
optical depth of the absorber at $j$, and $C$ is the effective covering fraction
of the absorber with respect to source $i$ at velocity bin $j$ (S04, G05).

The effective covering fraction at bin $j$
is the weighted sum of individual covering fractions summed over 
all emission sources (G05):
\begin{equation}
C_j = \sum_i C^i_j R^i_j
\end{equation}
G05 found that changes in the Ly$\alpha$ absorption profile
between the 2002 and 2003 epochs could be explained by the change in the
effective covering fraction arising from the change in the relative
contribution of the continuum and broad line region to the total flux
in each epoch.  No changes in the individual emission component covering fractions 
or in the \ion{H}{1} column density
are necessary to produce the observed reduction in depth of the absorption trough
in Components 2-2a between the 2002 and 2003 epochs.

\begin{figure}
\epsscale{0.8}
\plotone{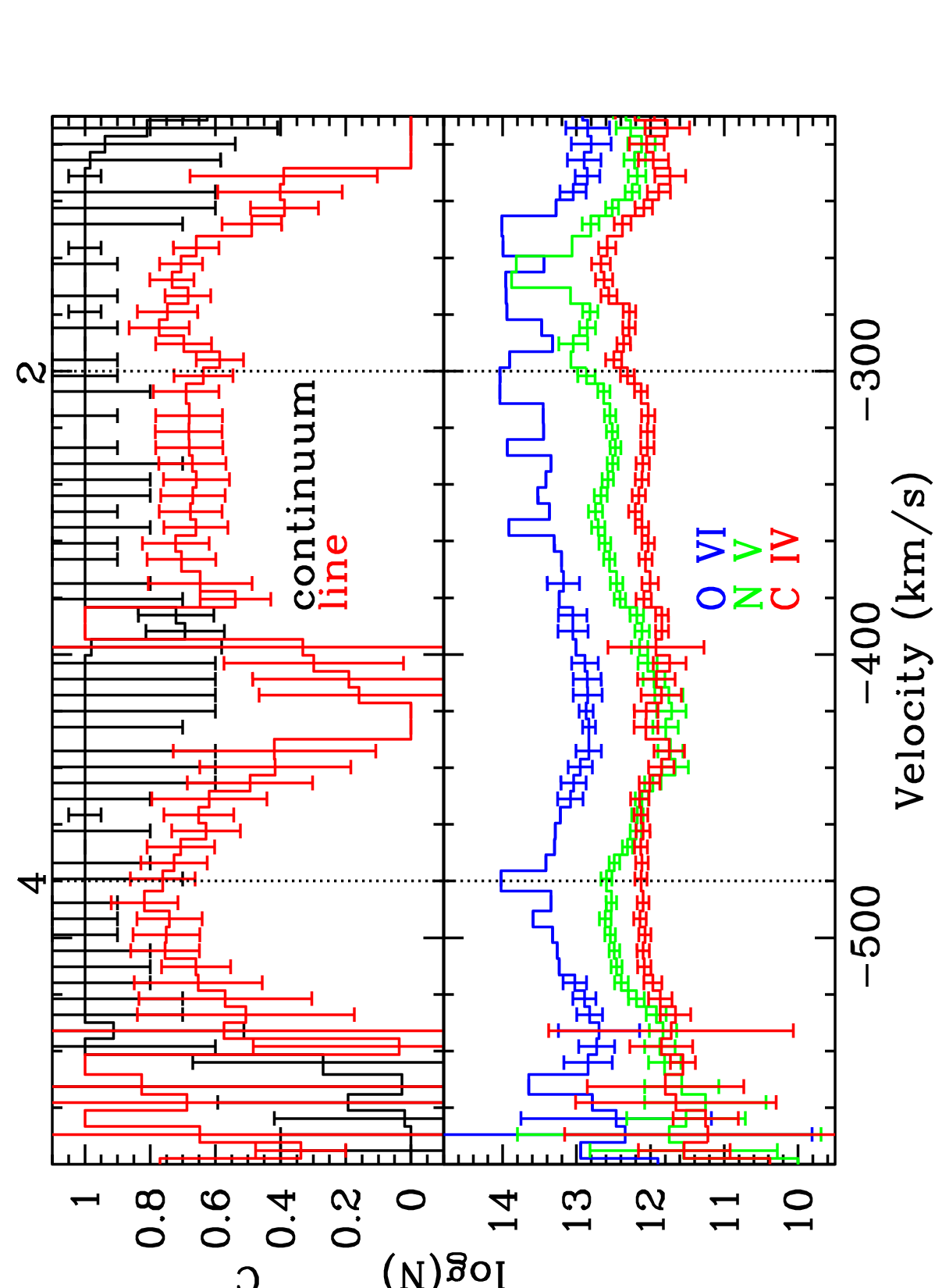}
\caption{
Global CNO doublet fit from Gabel et al.\ (2005)
to line and continuum covering fractions (top panel) and 
\ion{O}{6}, \ion{N}{5}, and \ion{C}{4} column densities (bottom panel).
Positions of Components 2 and 4 are
labeled at top of figure.
\label{fig:cno2003}}
\end{figure}

However, we found that this method cannot account for the changes in the absorption
profiles of \ion{O}{6}, \ion{N}{5}, and \ion{C}{4} between the 2002 and 2003
epochs if only the continuum and
emission line sources are considered separately.
Below we will examine this ionic absorption variability 
while treating the intermediate
width emission line components as a separate emission source.   
For \ion{O}{6}, we also include the 1999 epoch for which we have additional \fuse\ data.
We will use the well-constrained covering fraction and column density solutions
derived by G05 from the high flux state 2003 data shown in Figure~\ref{fig:cno2003}.
These solutions are the global CNO fits, the simultaneous fits to the
\ion{O}{6}, \ion{N}{5}, and \ion{C}{4} doublets described by G05.
The continuum covering fraction is effectively unity over the entire velocity
range of the absorption while the total emission line covering fraction varies.
For the remainder of this paper, however, we shall set 
$C^{\rm cont}=1$ (see also Arav et al.\ 2007).

First, however, in the following section
we will describe the general observed variation of the intrinsic absorption.
For each species, in each epoch, we show and discuss the normalized absorption
profiles in the following figures and sections
under the assumption
$C^{\rm ILR}=C^{\rm BLR}=C^{\rm cont}=1$.
We know this
assumption is not strictly true, given the covering
fraction and column density solutions shown in Figure~\ref{fig:cno2003}; but we
use this as a heuristic model and a starting point for further analysis.
In Section~\ref{sec-photo}, we discuss photoionization models for this outflow, 
and in Section~\ref{sec-cvsp} we discuss the importance of these 
photoionization effects relative to changes in the effective covering fraction
of the absorption resulting from variations in the underlying emission sources,
primarily the broad and intermediate velocity width emission lines.

\subsection{Description of Normalized Absorption Profiles}
\subsubsection{\ion{O}{6}}
\begin{figure}
\epsscale{0.8}
\plotone{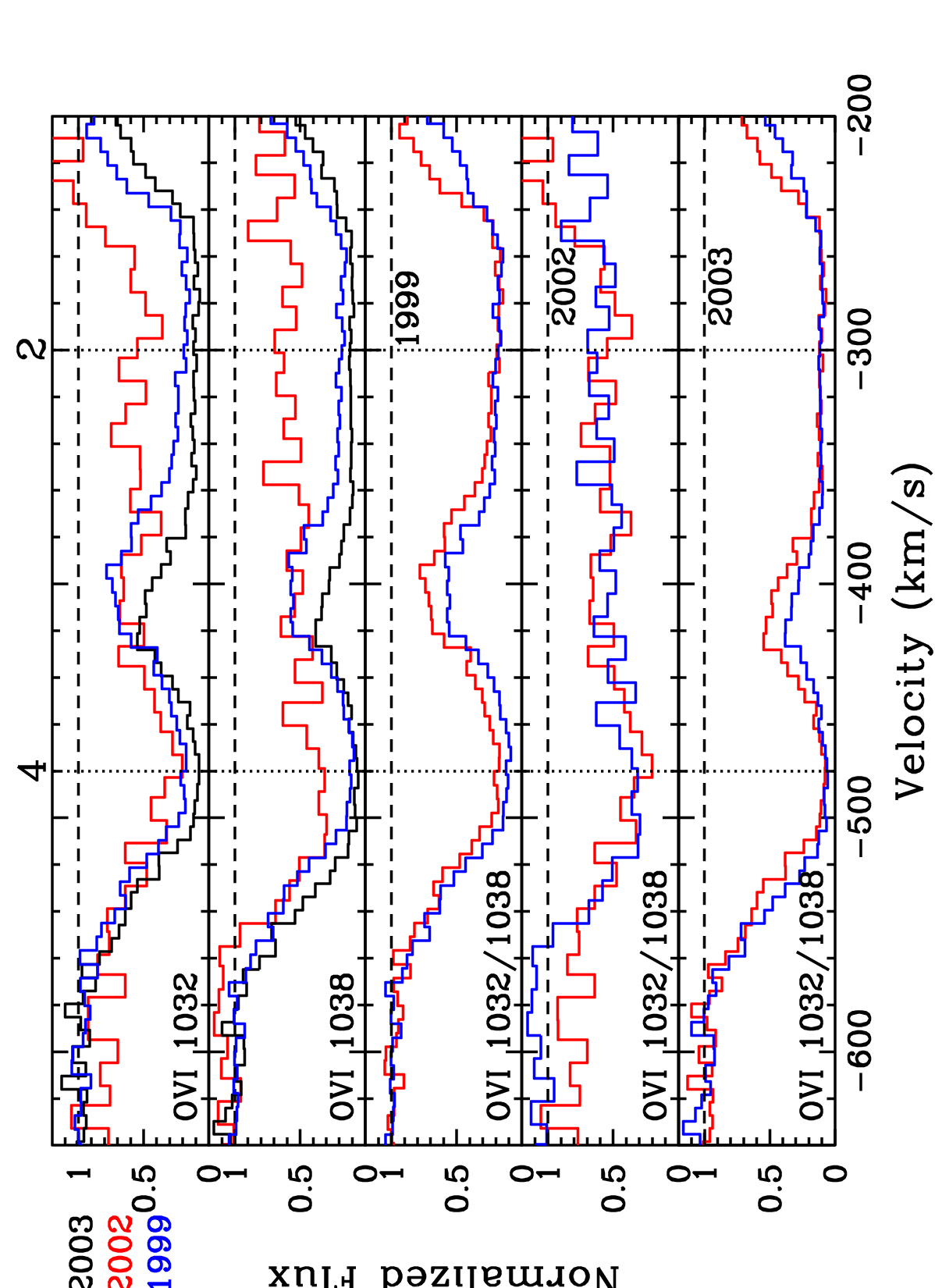}
\caption{Normalized \ion{O}{6} doublet over all observation epochs for 
$C_{\rm ILR}= C_{\rm BLR}=C_{\rm cont}=1$. Positions of Components 2 and 4 are 
labeled at top.  Top panel shows the blue doublet components
in the 1999, 2002, and 2003 epochs together, while the second panel shows this
for the red components.  Bottom three panels show the blue and red doublet components 
together in each of the three individual observation epochs.
\label{fig:o6doub}}
\end{figure}

Figure~\ref{fig:o6doub} 
shows the \ion{O}{6} profile.
Given the equal depths of the red and blue doublet components, it is clear that
the \ion{O}{6} absorption in the core of Component 4 is completely saturated
in the 2002 and 2003 epochs, and less so in the 1999 high flux state.
If $C^{\rm ILR}=1$ as assumed here,
the increase in absorption depth of the saturated core between
2002 and 2003 indicates the total 
effective covering fraction increased.
This is a contradiction to our assumption here, i.e. unity covering
fractions for all emission components for all epochs, confirming
what we know from Figure~\ref{fig:cno2003}, namely that
$C^{\rm BLR}$ and possibly also $C^{\rm ILR}$ are not unity.

The \ion{O}{6} absorption is saturated in Component 2 in all three
epochs.  For $C^{\rm ILR}=1$, the top panel of Figure~\ref{fig:o6doub} shows that
the absorption depth is largest
in 2003 and smallest in 2002.  This is true of Component 4 as well. 
These are the epochs of the
smallest and largest ILR/BLR flux, respectively, indicating 
that the profile change is caused by the change in this emission line ratio,
contributing to a change in the total effective covering fraction.

\subsubsection{\ion{N}{5}}
\begin{figure}
\epsscale{0.8}
\plotone{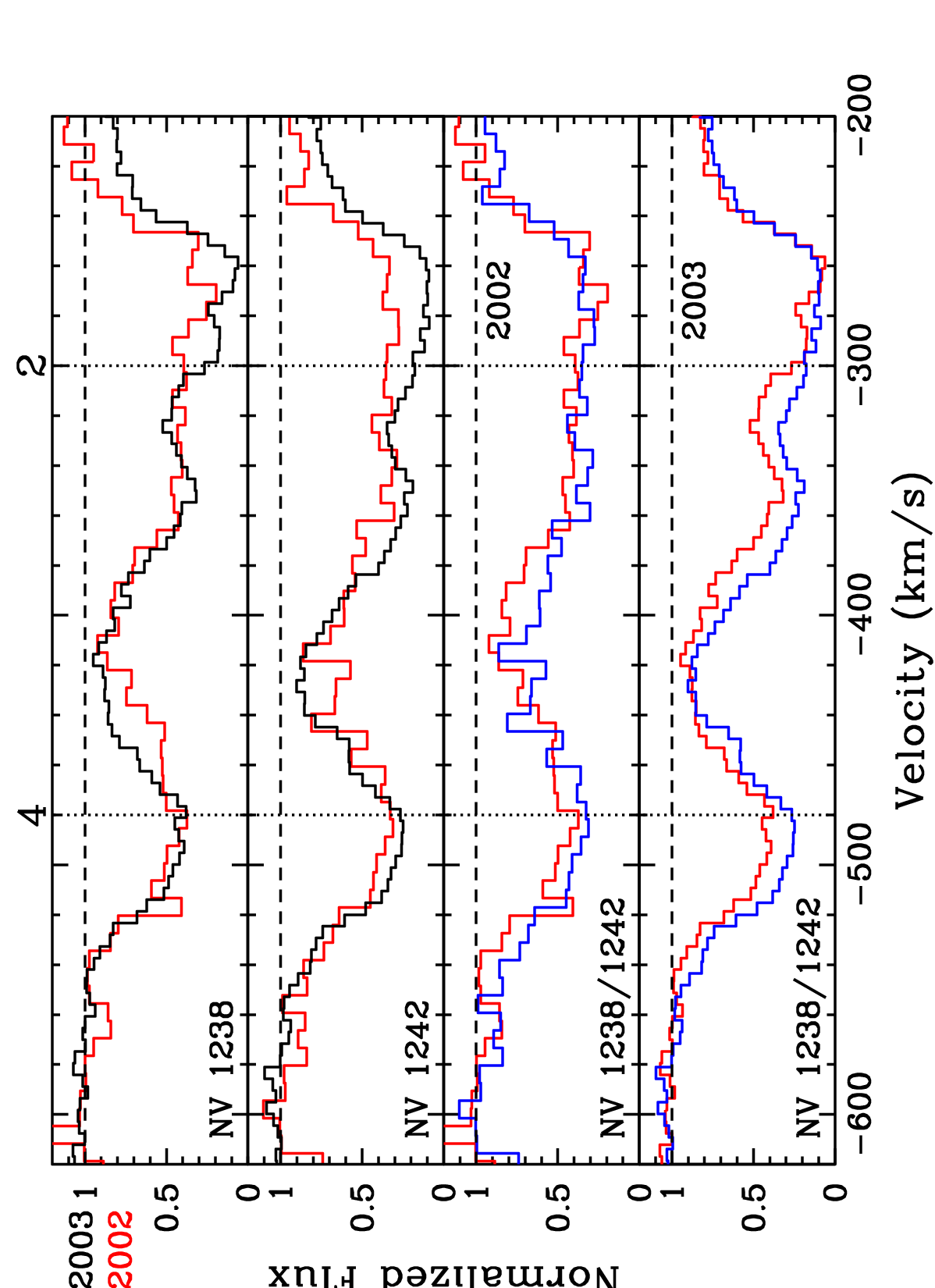}
\caption{Normalized \ion{N}{5} doublet over all observation epochs for 
$C_{\rm ILR}= C_{\rm BLR}=C_{\rm cont}=1$. Positions of Components 2 and 4 are
labeled at top.  Top panel shows the blue doublet components
in both the 2002 and 2003 epochs together, while the second panel shows this
for the red components.  Bottom three panels show the blue and red doublet components
together in each individual observation epoch.
\label{fig:n5doub}}
\end{figure}

In Component 4, the entire \ion{N}{5} profile is marginally saturated in 2002
and unsaturated in 2003.  For $C^{\rm cont}=C^{\rm BLR}=C^{\rm ILR}=1$,  
as shown in Figure~\ref{fig:n5doub},
the absorption 
depth is similar but nonzero in both epochs. 
This indicates once again that the emission line covering fractions are
nonunity, but also that the change in the effective line covering fraction 
between the two epochs is not as dramatic as for \ion{O}{6}.

For Component 2, the \ion{N}{5} profile is
saturated in 2002, but in 2003 it is saturated in the core and in the red
wing of the profile, while the blue wing is unsaturated.
The top panels of Figure~\ref{fig:n5doub} show
the absorption depth in 2003 is greater than in 2002,
particularly in the consistently saturated regions.
This could be the result of the changing ILR/BLR flux ratio and the
resulting variation in the effective line covering fraction, as described above
for the \ion{O}{6} profile. 

\subsubsection{\ion{C}{4}}
In 2002, Component 4 of the \ion{C}{4} profile is possibly saturated in the
red wing and unsaturated elsewhere, while in 2003
it is unsaturated over most of the profile.
The absoprtion depth decreases from 2002 to 2003, indicating
that the \ion{C}{4} column density decreased as the continuum
flux increased by a factor of $\sim8$ between 2002 and 2003.
The profile depth change between 2002 and 2003
is less than expected for a simple column density response
to this flux increase, leading to the
conclusion that the total effective covering fraction decreased over this time.
Although there is only a small change in the ILR/BLR emission flux ratio 
(Fig.~\ref{fig:c4_em}) the continuum level relative to the emission lines rises
significantly between 2002 and 2003. 
The effective covering fraction thus rises since $C^{\rm cont}=1$.

In Component 2, the \ion{C}{4} profile is unsaturated in both 2002 and 2003, and
the absorption depth is unchanged between these two epochs, most likely also due to
ionization changes balancing an increase in the effective covering fraction.

\subsubsection{Lyman series}
\begin{figure}
\epsscale{0.8}
\plotone{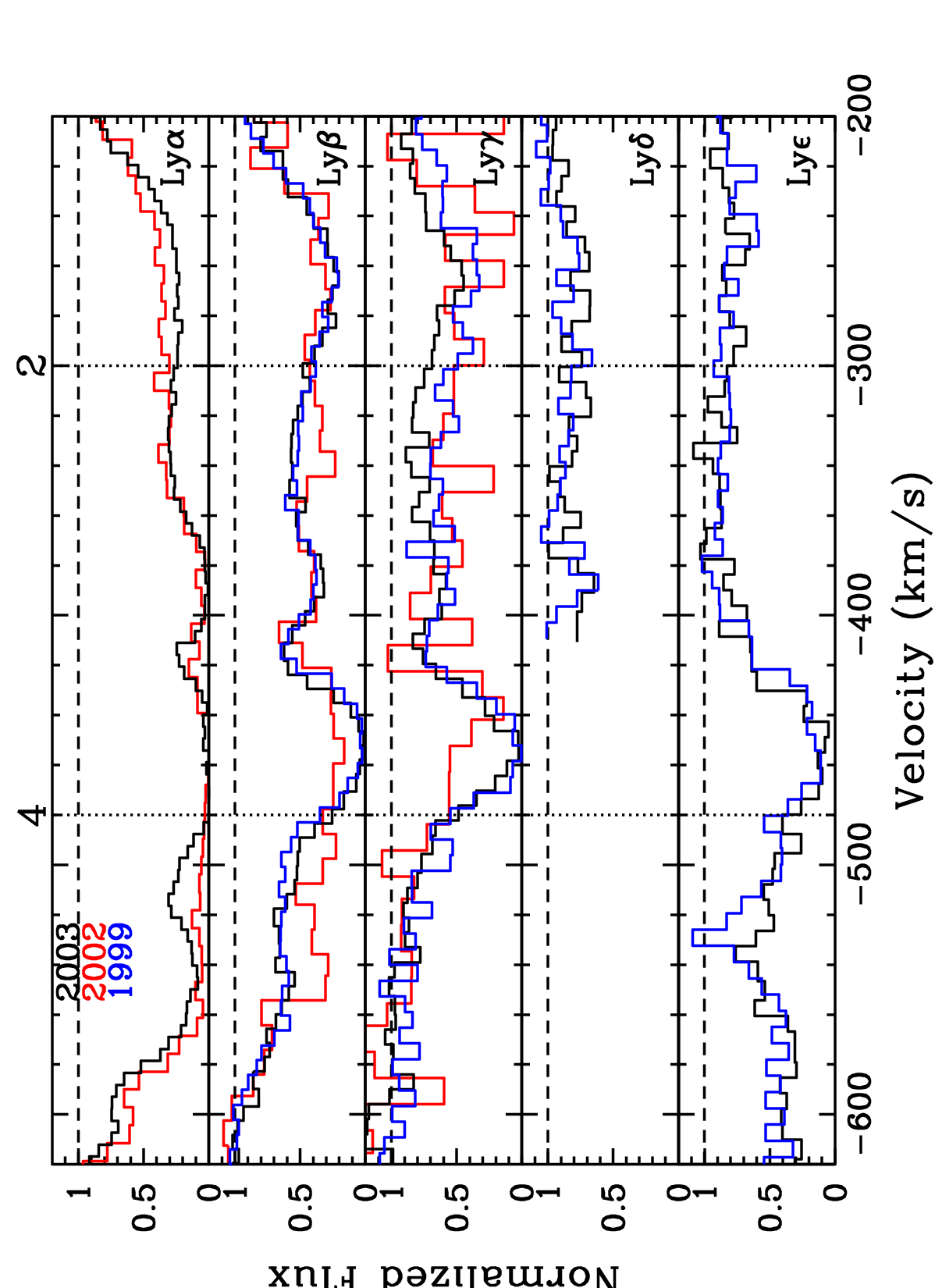}
\caption{Normalized Lyman series over all observation epochs for 
$C_{\rm ILR}= C_{\rm BLR}=C_{\rm cont}=1$. Positions of Components 2 and 4 are
labeled at top. Each panel shows an individual component of the Lyman series in all
observation epochs:  2002 and 2003 only for Ly$\alpha$; 1999, 2002, and 2003 for
Ly$\beta$ and Ly$\gamma$; and 1999 and 2003 only for Ly$\delta$ and Ly$\epsilon$.
\label{fig:lyseries}}
\end{figure}

For \ion{H}{1}, the only part of the absorption complex that can
be reliably attributed to intrinsic gas is Component 2 
($v \approx 200-300$ km s$^{-1}$).
In this velocity component, the \lya\ absorption depth
is noticeably weaker in 2002 than in 2003.  As pointed out
by G05, this is explained by a lower effective
covering fraction in 2002 due to a larger relative contribution
by the BLR relative to the continuum and the fact that $C^{\rm cont}=1$ while 
$C^{\rm BLR}<1$.   
For our assumption $C^{\rm ILR}=C^{\rm BLR}=1$, shown in Figure~\ref{fig:lyseries}, 
we see this reported
change in the \lya\ profile, and little change in the Ly$\beta$ profile, which we 
would expect to be more influenced by the changes in the \ion{O}{6} emission components.

For other lines further along the Lyman series from
Ly$\beta$, 
there is little or no underlying 
emission line flux beneath these profiles.
Examining Ly$\gamma$, a Lyman series line unlikely to be affected by saturation,
we find no discernable difference in the profiles
from 1999 to 2002, but the profile does appear less deep in 2003.
However there is no difference in  Ly$\delta$ or Ly$\epsilon$ between
the two high flux states, 1999 and 2003.

\section{Photoionization}
\label{sec-photo}
\begin{figure}
\epsscale{0.9}
\plotone{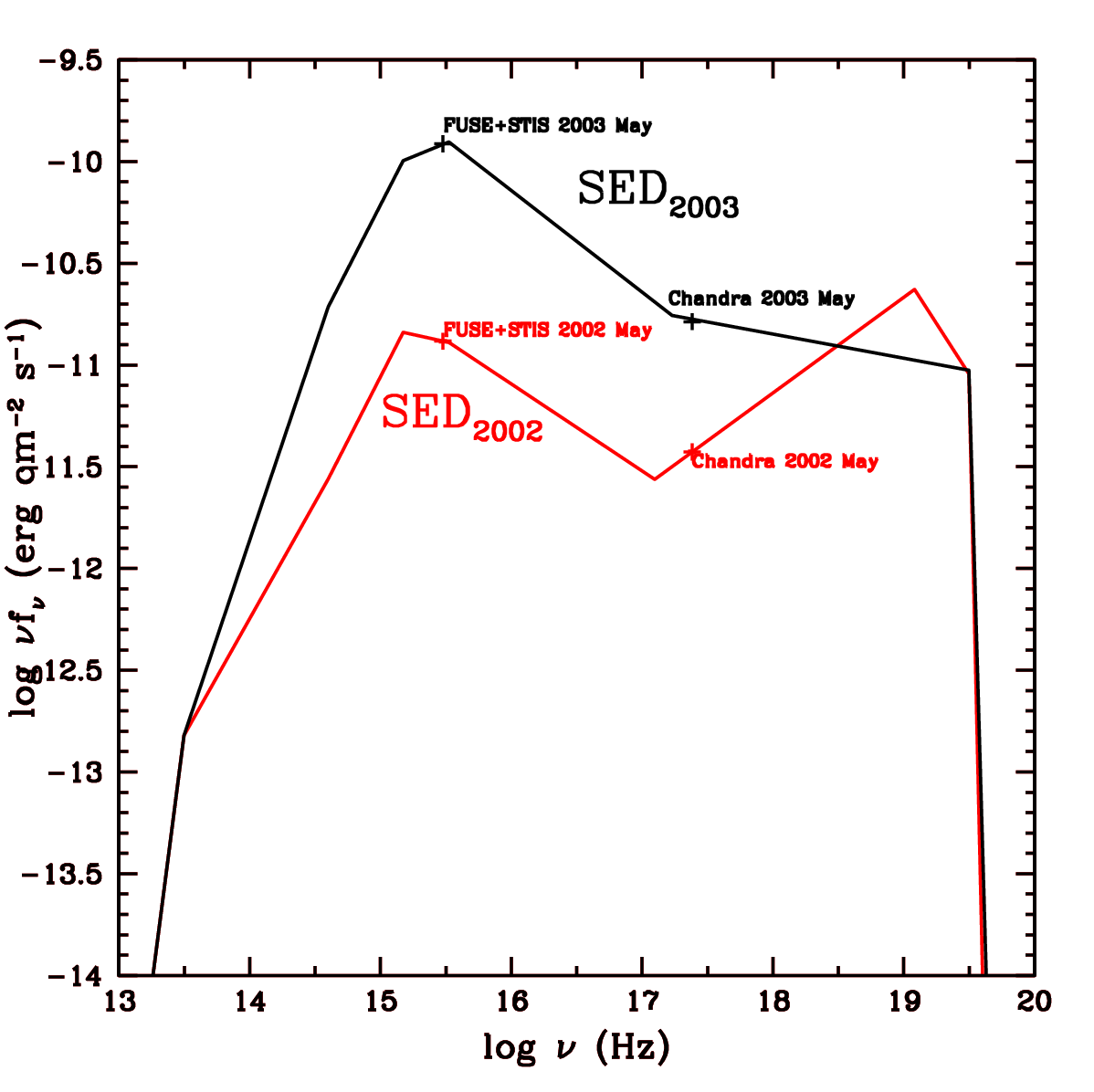}
\caption{Spectral energy distribution of Mrk 279 in 2002 and 2003, based
on \fuse\ and \chand\ observations.
\label{fig:sed}}
\end{figure}

To investigate a wider range of possiblities for the various emission source covering
fractions, and to
disentangle the effects of photoionization from those of covering fraction and emission
source variation on the various species considered above,
we computed photoionzation models similar to those of 
Krolik \& Kriss 
(1995, 2001) using the XSTAR code (Kallman 2000).
We calculate the photoionization state 
of the gas using the supersolar abundance found by Arav et al.\ (2007) 
spectral energy distributions (SEDs) appropriate to the 2002 and 2003
\fuse\ and \chand\
observations shown in Figure~\ref{fig:sed}.
Both SEDs are normalized to the measured \fuse\ and \chand\
spectra over the spectral range of these instruments. The parameters of the \fuse\ spectra are
discussed in
Section~\ref{sec-cont}.   
We connect the 2002 \fuse\ and \chand\ measurements with 
$\alpha_{\rm UV-X-ray}=0.57$; and we
approximate the 2002 \chand\ double power law spectrum (S04) 
with a single power law with $\alpha=1.47$ from 0.5 to 50 keV and we extend the SED to $\sim$150 keV with
$\alpha=0$.
We approximate the 2003 \chand\ power law + modified blackbody continuum (Costantini et al.\ 2007)
with a smooth power law: $\alpha =0.5$ connecting the \fuse\ and \chand\ data to 0.7 keV
and $\alpha=0.88$ 
to $\sim$150 keV (Costantini et al.\ 2007, Arav et al.\ 2007).  
For energies lower than the \stis\ bandpass and higher than the \chand\ bandpass, we use
cut-offs for a standard
AGN spectrum (cf.\ Costantini et al.\ 2007).
We executed the model with grids in input column density and ionization parameter
from $\log N = 18$ to $\log N = 21$ 
and from $U=0.01$ to $U=10$, respectively.

In Figure~\ref{fig:ratio}, we show results from the $\log N = 20$ run.
We plot the ratio of the column densities of 
\ion{C}{3}, \ion{C}{4}, \ion{N}{5}, and \ion{O}{6} in the XSTAR simulations
using SED$_{2002}$ to the values from the XSTAR simulations using SED$_{2003}$ 
as a function of the ionization parameter in the SED$_{2002}$ simulations.  
To take this ratio, we compare the column densities in each simulation at
equivalent ionization parameters given the different SEDs.  The number of 
ionizing photons supplied by SED$_{2002}$ is about one-tenth the number 
supplied by SED$_{2003}$.  Therefore, we take the ratio of the column
densities of each ion at $U_{2002}$ and $U_{2003}=10U_{2002}$.  These
are plotted against $U_{2002}$ in Figure~\ref{fig:ratio}.

\section{Effective Covering Fraction versus Photoionization}
\label{sec-cvsp}
\begin{figure}
\epsscale{0.9}
\plotone{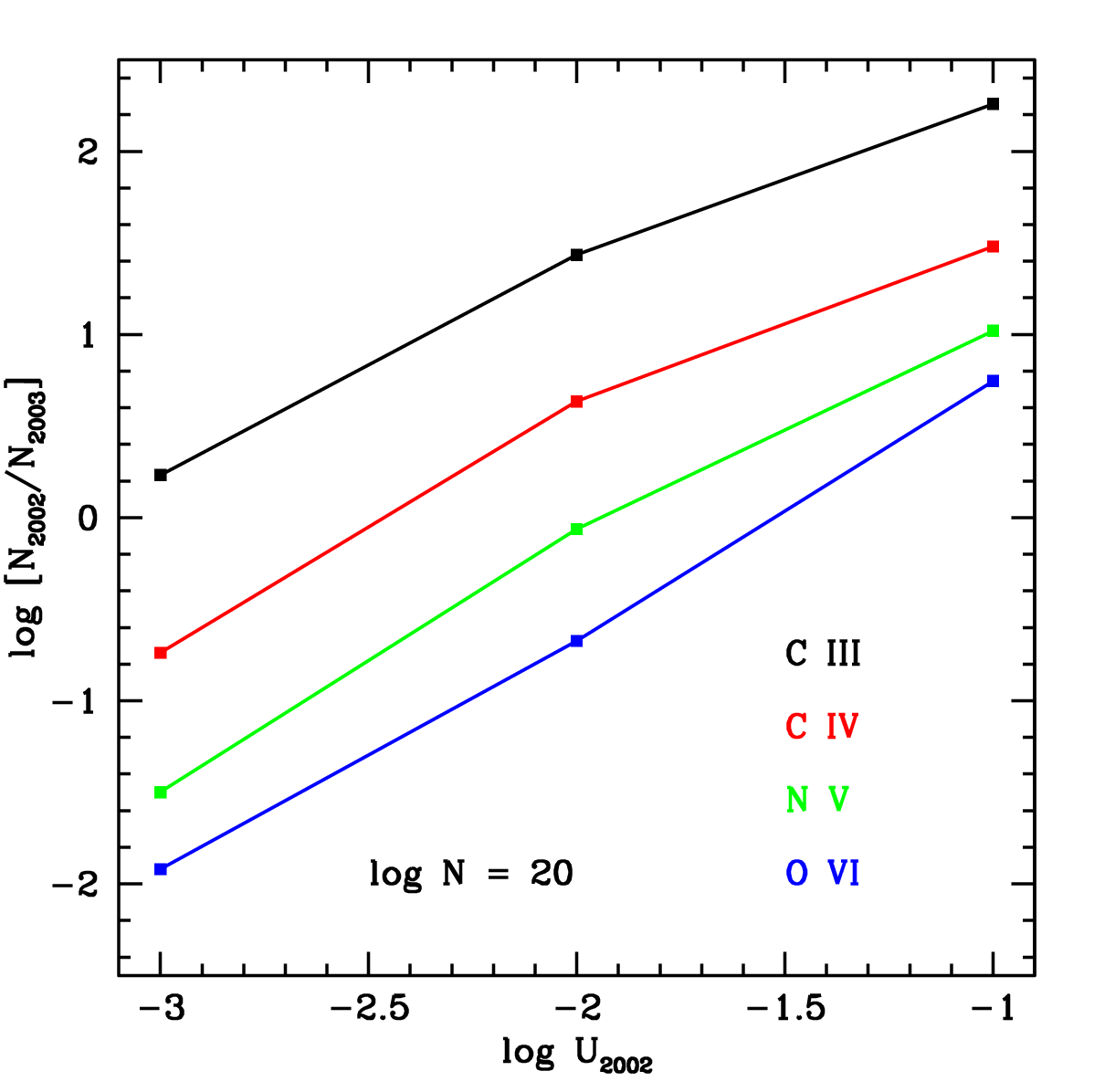}
\caption{
Ratio of the column density of 
\ion{C}{3}, \ion{C}{4}, \ion{N}{5}, and \ion{O}{6} in the XSTAR 
simulations using SED$_{2002}$ to those using SED$_{2003}$ as a function of the 2002 ionization paramter.
The total input column density for all is $\log N = 20$.
\label{fig:ratio}}
\end{figure}

We now seek to establish whether
the changes in the absorption profiles of the 
highly ionized species \ion{C}{4}, \ion{N}{5}, and \ion{O}{6}
between
the 2002 and 2003 epochs can be explained by partial covering of some portion of the 
line emission combined with changes in the relative contributions of the component broad and
intermediate line
fluxes to the total emission profile, or if some change in the column density in response
to changes in the continuum flux is required by the data.
To address this, we use the well-constrained 2003 covering fractions and column
densities derived by G05 and attempt to reproduce the 2002 absorption profiles
in two ways:  (1) by allowing the covering fraction of the ILR to be different from
that of the BLR; and (2) by varying the column densities according to changes in the
ionizing flux as predicted by our photoionization models.
From Figure~\ref{fig:ratio}, we see how we expect the column densities of
each ion to change under photoionization alone.

From the emission line fits, we consider three emission sources,
the continuum, the BLR, and the ILR. Thus, 
Equation~\ref{equ:int} becomes
\begin{equation}
I_j =  R^{\rm cont}_j (C^{\rm cont}_j e^{-\tau_j} + 1 - C^{\rm cont}_j) +
R^{\rm BLR}_j (C^{\rm BLR}_j e^{-\tau_j} + 1 - C^{\rm BLR}_j) +
R^{\rm ILR}_j (C^{\rm ILR}_j e^{-\tau_j} + 1 - C^{\rm ILR}_j).
\end{equation}
We assume $C^{\rm cont}=1$, consistent with G05. 
For doublets, however, the resulting system of equations is still underconstrained.  
Therefore,
we assume three possibilities for $C^{\rm ILR}$:
(1) $C^{\rm ILR}=C^{\rm BLR}$; (2) $C^{\rm ILR}=0.5 C^{\rm BLR}$; and 
(3) $C^{\rm ILR}=0$.

The overall goal of this experiment is to reproduce the 2002 spectrum from
the column densities derived from the higher quality 2003 data.   We
attempt to do this by varying both the changes in the ionic column densities
using our photoionization modeling as a guide, and by varying the covering fractions
of the intermediate line region using the three cases above.
Here is a summary of our method:
\begin{itemize}

\item{We begin with the well-constrained solutions for the
the emission line covering fractions from the 2003 data presented by G05.
In that solution, the line covering fractions of \ion{C}{4}, \ion{N}{5}, and
\ion{O}{6} were tied to the same value, 0.64 and 0.76 for the centers of Components 2 and 4, 
respectively.
}

\item{We use
the new emission line fits to the 2003 data presented in Section~\ref{sec-cont}, 
including intermediate line components,
to derive $C^{\rm BLR}$ and $C^{\rm ILR}$ from the values of $R^{\rm BLR}$ and $R^{\rm ILR}$ at each
velocity bin, from the total line covering fraction given by G05.
Note that in doing this, we are formally calculating {\it different} covering fractions
for \ion{C}{4}, \ion{N}{5}, and \ion{O}{6} because the emission line flux underlying
each set of doublets is different. We will return to this point in the discussion in 
Section~\ref{sec-disc}.}

\item{We 
apply these to the 2002 epoch given the values of $R^{\rm BLR}$ and $R^{\rm ILR}$ derived 
from that spectrum to examine how the profiles may have changed from changes in these
ratios alone.}

\item{We also examine how the profiles change if the column densities
of the various ions have changed as well, by reasonable factors gleaned from our
photoionization modeling as shown in Figure~\ref{fig:ratio}.}

\item{We center the x-axis of 
Figure~\ref{fig:ratio} and concentrate our attention on the case $\log(U_{2003})\sim-1$
(and $\log(U_{2002})\sim-2$)
since this is the result of the detailed photoionization modeling and abundance analysis
of Arav et al.\ (2007).}

\end{itemize}

\section{Discussion}
\label{sec-disc}
\begin{figure}
\epsscale{0.8}
\plotone{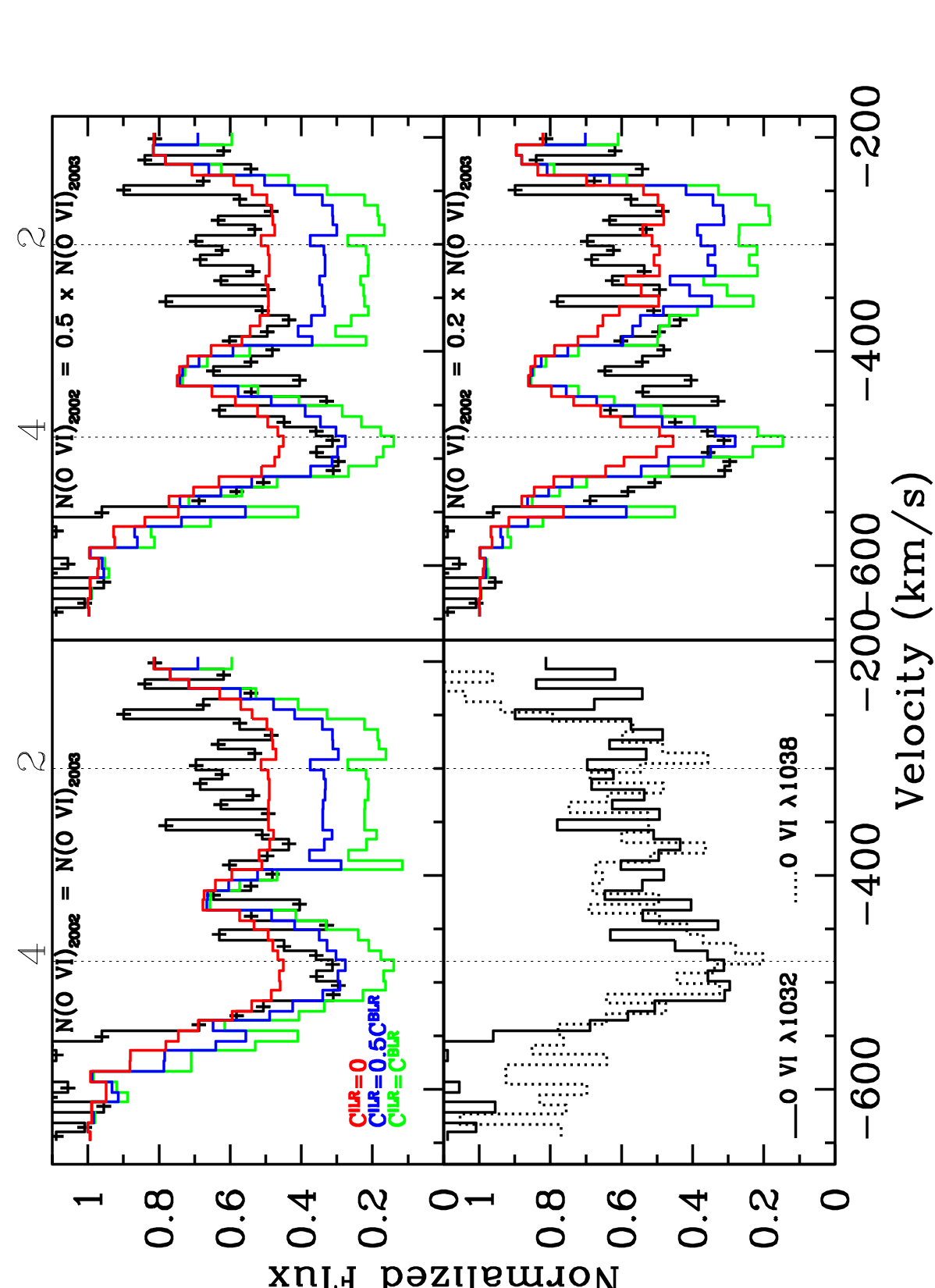}
\caption{Blue and red components of \ion{O}{6} profile from 2002 \fuse\ data (lower left); blue component
is repeated in other windows with profiles expected from \ion{O}{6} column density measured from 2003 \fuse\
data and different ILR covering fractions:  $C_{\rm ILR}=0$ (red), $C_{\rm ILR}=0.5C_{\rm BLR}$ (blue), 
$C_{\rm ILR}=C_{\rm BLR}$ (green). Positions of Components 2 and 4 are
labeled at top.
\label{fig:o6land}}
\end{figure}

\begin{figure}
\epsscale{0.8}
\plotone{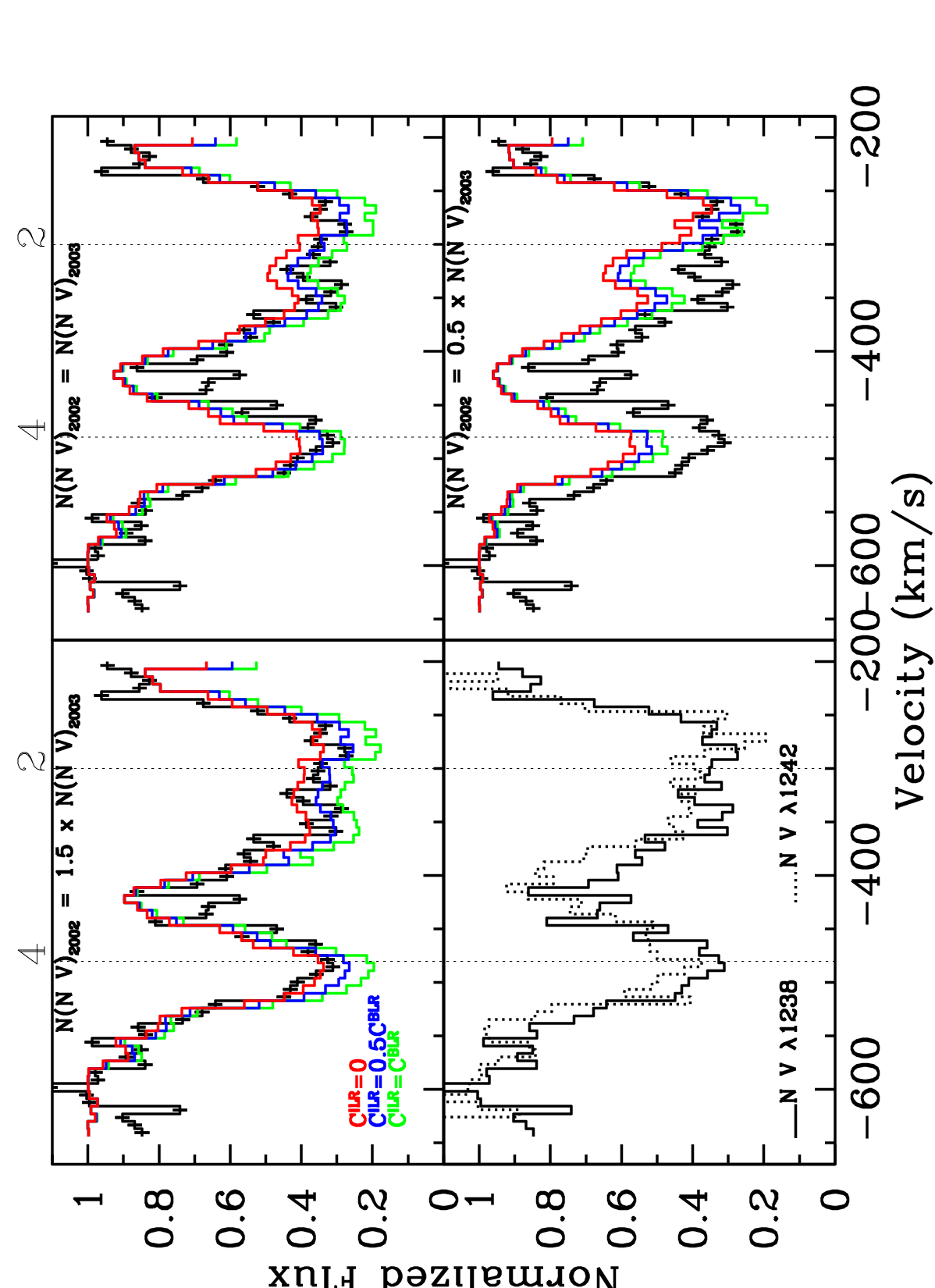}
\caption{Blue and red components of \ion{N}{5} profile from 2002 \stis\ data (lower left); blue component
is repeated in other windows with profiles expected from \ion{N}{5} column density measured from 2003 \stis\
data and different ILR covering fractions:  $C_{\rm ILR}=0$ (red), $C_{\rm ILR}=0.5C_{\rm BLR}$ (blue), $C_{\rm ILR}=C_{\rm BLR}$ (green). Positions of Components 2 and 4 are
labeled at top.
\label{fig:n5land}}
\end{figure}

\begin{figure}
\epsscale{0.8}
\plotone{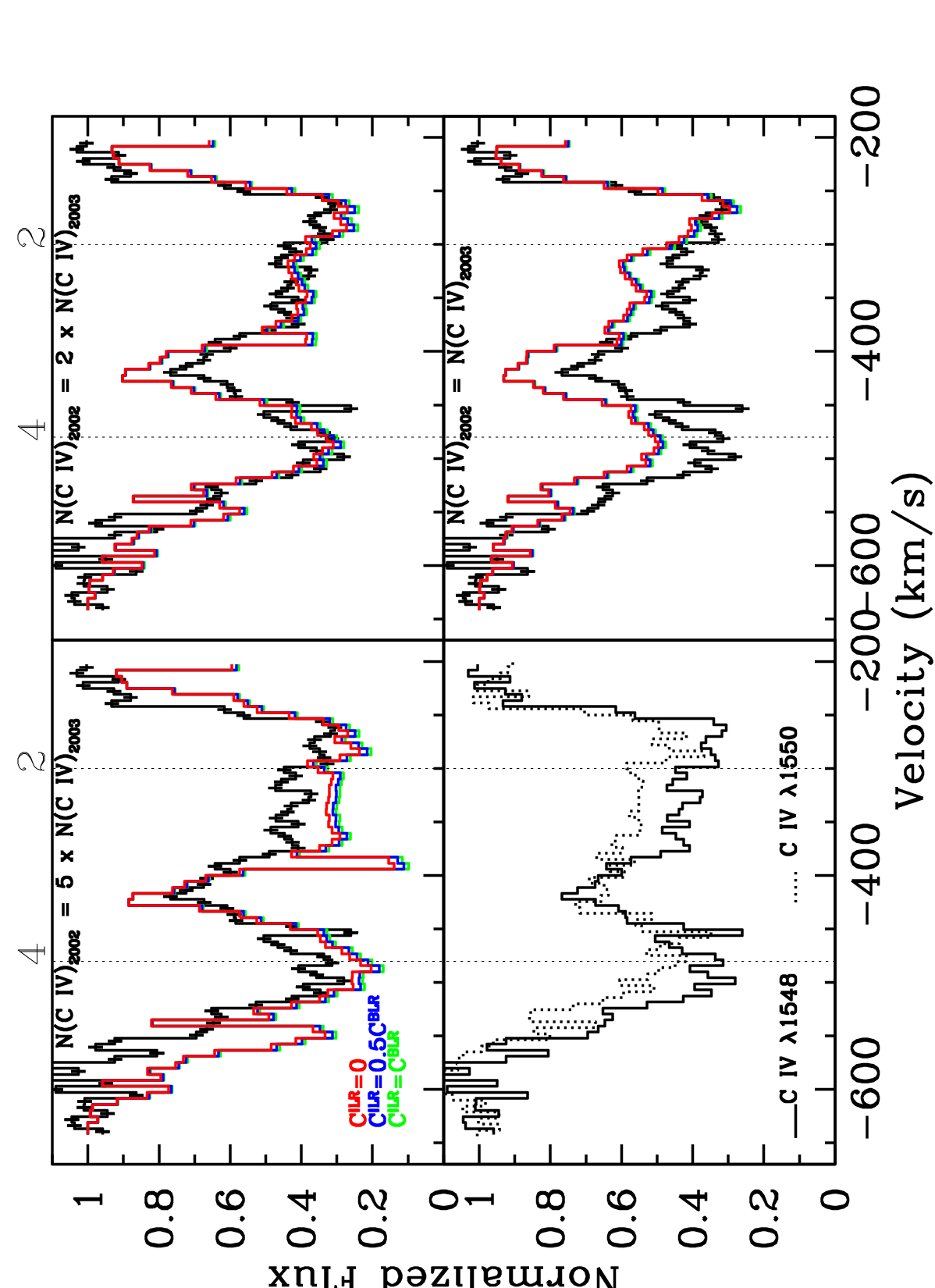}
\caption{Blue and red components of \ion{C}{4} profile from 2002 \stis\ data (lower left); blue component
is repeated in other windows with profiles expected from \ion{C}{4} column density measured from 2003 \stis\
data and different ILR covering fractions:  $C_{\rm ILR}=0$ (red), $C_{\rm ILR}=0.5C_{\rm BLR}$ (blue), $C_{\rm ILR}=C_{\rm BLR}$ (green). Positions of Components 2 and 4 are
labeled at top.
\label{fig:c4land}}
\end{figure}

The results for \ion{O}{6}, \ion{N}{5}, and \ion{C}{4} 
are shown in Figures~\ref{fig:o6land}-\ref{fig:c4land}.  In each lower left panel, we show
the blue and red components of the doublet profiles from the 2002 data. 
In the other three quadrants, we
show the blue component of the 2002 profile in black.  Superimposed on this are the attempts
to reproduce this profile from the  parameter constraints from the high
signal-to-noise 2003 data described above,
under variations in both the column densities
and the covering fractions.
The column density changes from 2002 to 2003 increase
clockwise from top left to bottom right, with one panel
for each ion showing how the profiles would change if {\it no} change in the ion
column density occured between 2002 and 2003. 

Figure~\ref{fig:o6land} shows that because the
\ion{O}{6} profile is saturated, the column density 
is unconstrained. For Component 2, regardless of the column density the only value
of $C^{\rm ILR}$ consistent with the profile is zero, while for Component 4,
$C^{\rm ILR}=0.5 C^{\rm BLR}$.  The photoionization model predicts an increase
of about a factor of four in the \ion{O}{6} column density from 2002 to 2003,
given the change in the continuum level over that period.

For \ion{N}{5}, shown in Figure~\ref{fig:n5land}, 
the best fit solution for both
Components 2 and 4 is
a column density decrease of 
$\sim$30\% from 2002 to 2003 and 
$C^{\rm ILR}=0$, although $C^{\rm ILR}=0.5C^{\rm BLR}$ also provides an acceptable
fit for Component 4.   
Our pure photoionization model would predict
a decrease of about one half that amount in the \ion{N}{5} column density between 2002 and 2003.

For \ion{C}{4}, Figure~\ref{fig:c4land} shows 
that a decrease in the column density from 2002 to 2003 is required to 
match the profiles in both components, regardless of the value of
$C^{\rm ILR}$.  A change in the column density by larger than a factor of
$\sim$2, however, is not consistent with the observed absorption depths. 
Our photoionization modeling indicates that we would expect
a larger decrease in the \ion{C}{4} column
density from photoionization alone, a factor of nearly six.
This covering fraction is relatively poorly constrained because
the relative contributions of the BLR and ILR change very little
between the 2002 and 2003 epochs, as noted above.
Of our models, the best fits to the 2002 spectrum come from 
$N$(\ion{C}{4})$_{2002}$=$2 \times N$(\ion{C}{4})$_{2003}$
and $C^{\rm ILR}=0$ for Component 2 and $C^{\rm ILR}=C^{\rm BLR}$ for
Component 4.   This ILR covering fraction solution for Component 4, however, is
not significantly different from $C^{\rm ILR}=0.5C^{\rm BLR}$.  

In the formal solution for these covering fractions we assumed
that the effective line covering fractions
for \ion{O}{6}, \ion{N}{5}, and \ion{C}{4} are all the same in the 2003 epoch
but that the resulting individual BLR and ILR covering
fractions are not necessarily equal, and that those individual BLR and ILR covering
fractions did not change between 2002 and 2003.
However, from the success of the G05 global CNO covering fraction fits, 
the covering fractions for each of these highly ionized species are likely to be comparable.
Thus, we summarize the results detailed above in this context and find that 
(1) the BLR covering fraction of \ion{C}{4}, \ion{N}{5}, and \ion{O}{6}
in the cores of Components 2 and 4 are $\sim$0.8-0.9; 
(2) the  ILR covering fractions
are zero for Component 2 and $\sim$0.4 for Component 4; and
(3) the column density in \ion{O}{6} decreased by an indetermiate factor while
the column density in \ion{N}{5} decreased by one third, and
that of \ion{C}{4} decreased by a factor of $\sim$2
between the two epochs.
Should we choose to constrain
the individual broad and intermediate emission line covering fractions of
\ion{C}{4}, \ion{N}{5}, and \ion{O}{6} 
to be equal (like G05, but unlike S04), we could use this set of three
doublets to solve for six variables, the continuum, broad line, and emission line
covering fractions and the column densities of \ion{O}{6}, \ion{N}{5}, and \ion{C}{4} 
as a function of velocity across the absorption profiles.
We leave this for future work.

\begin{figure}
\epsscale{0.9}
\plotone{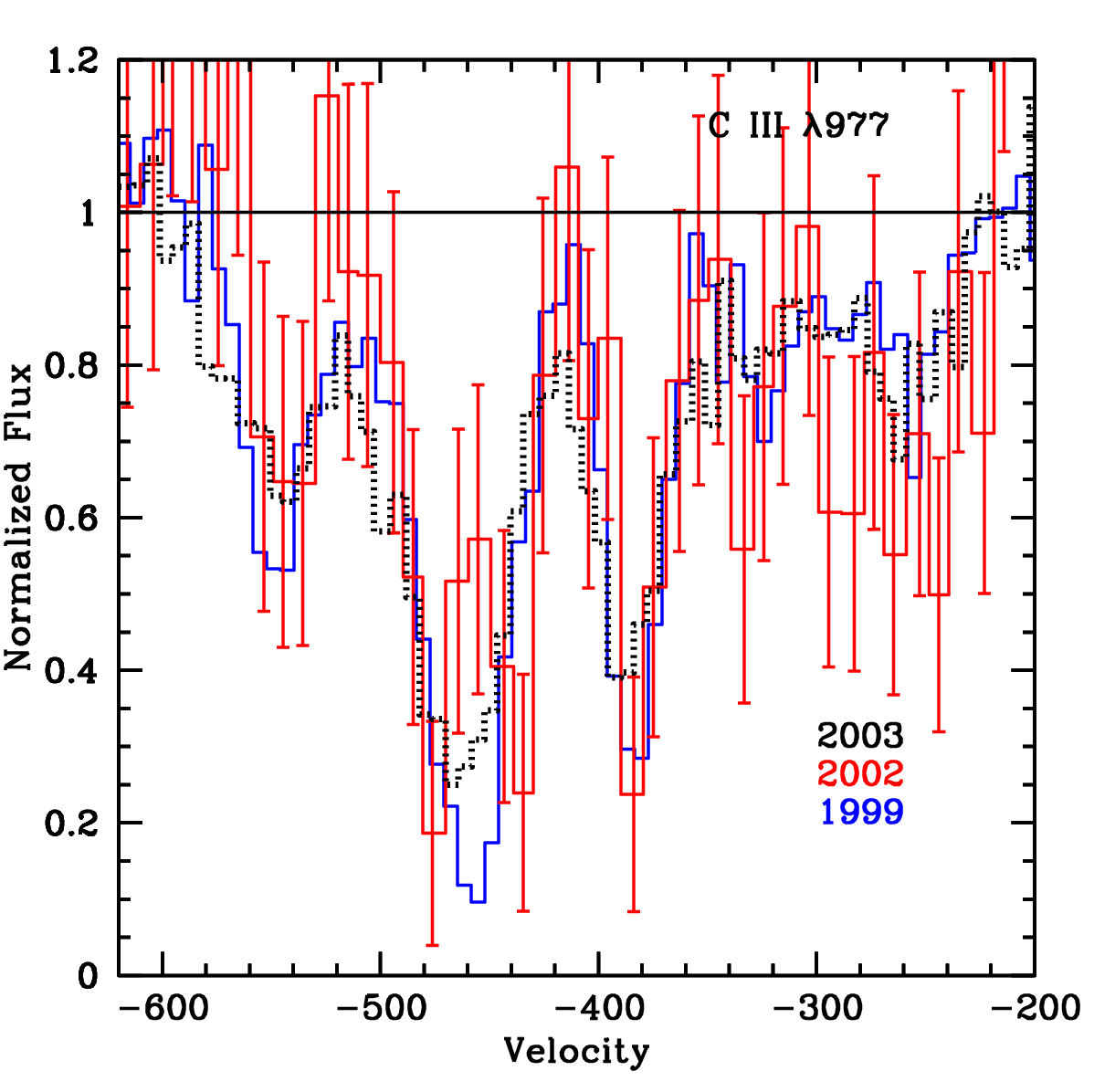}
\caption{The \ion{C}{3}~\lam977 profile in the 1999, 2002, and 2003 \fuse\ observations.
\label{fig:c3}}
\end{figure}

The \ion{C}{3} profile presents somewhat of a puzzle in light of the 
results we have outlined thusfar.  We expect the 
column density of this species to be quite sensitive to the factor of
$\sim$8 increase in the ionizing flux between the 2002 and 2003 observations.
Figure~\ref{fig:ratio} shows that for $U_{2002}=0.01$, 
the \ion{C}{3} column density should decrease by a factor of about 25-30 from 2002 to 2003.
In Figure~\ref{fig:c3}, we demonstrate that we see no significant change in the \ion{C}{3} profile
in either Component 4 or in Component 2.

One way to resolve this problem is to invoke a long photoionization response time for
the gas.
To test this possibility, we follow an equilibrium photoionization model of the gas
over time.  We began with a model that matches the well-constrained 2003 column densities
and dropped the photoionizing flux by a factor of 8 to evolve the absorber back in time to
the 2002 epoch. Following the trend of ionic abundances,
we find that the \ion{C}{3} column
density should decrease by a factor of two in about one hour for 
$n_{\rm e} = 3 \times 10^4$ cm$^{-3}$,
in about one day for $n_{\rm e} = 1500$ cm$^{-3}$ and in about one year for
$n_{\rm e} = 5$ cm$^{-3}$.  
Therefore, for a very small change in the \ion{C}{3} column density, a small total density
is required if the absorbing gas is in equilibrium.   
This density, along with $\log U = -1$ and the luminosity of Mrk~279, require that the gas lie at
$\sim$2 kpc from the ionizing source if it is part of the Mrk~279 outflow. 
This would certainly present us with a conundrum given the nonunity covering fractions
we have consistently found for the emission lines in absorption Components 2 and 4. 

\begin{figure}
\epsscale{0.9}
\plotone{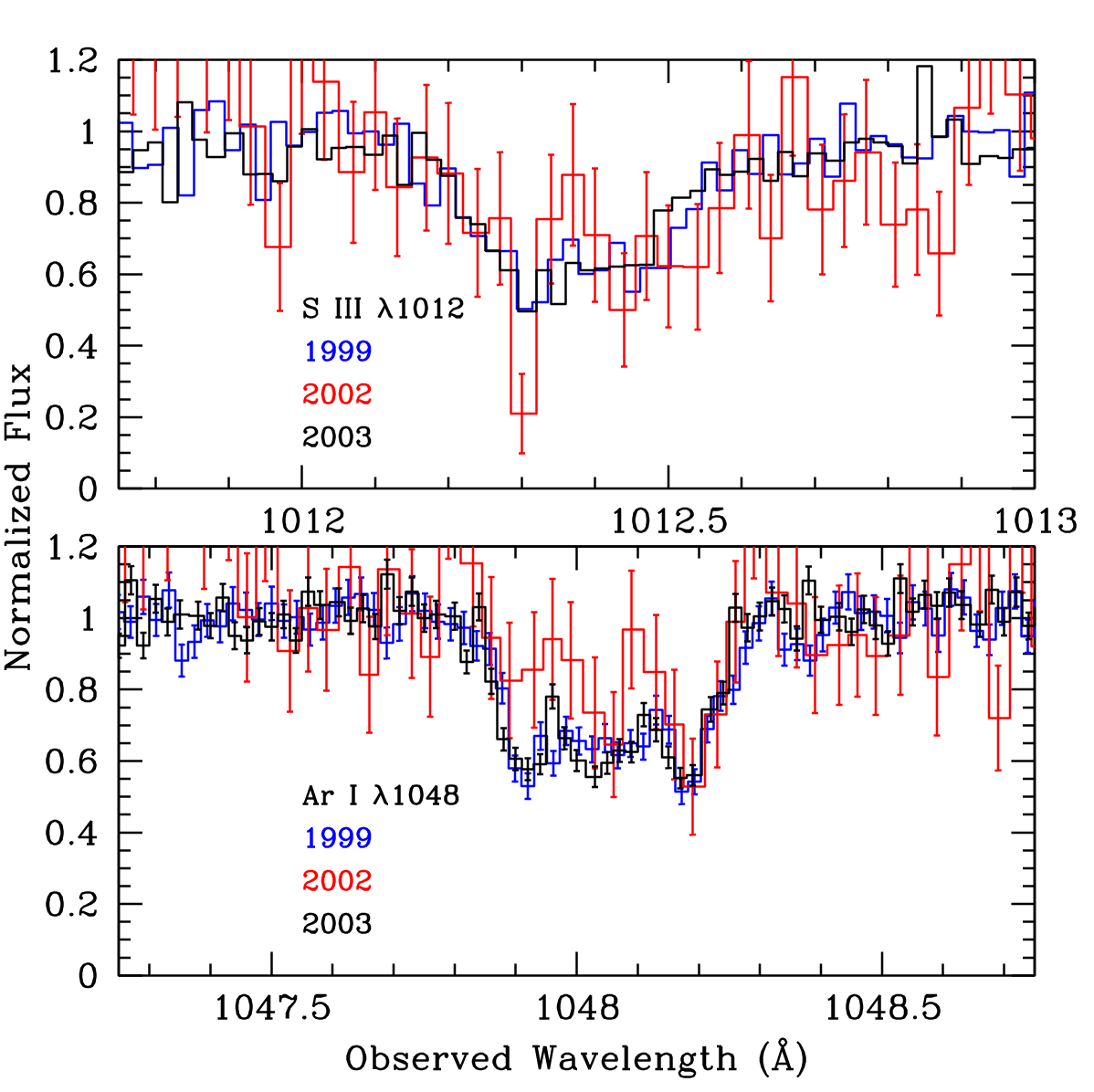}
\caption{Absorption profiles of interstellar species \ion{S}{3} and \ion{Ar}{1} in the
1999, 2002, and 2003 epochs of \fuse\ data.
\label{fig:ism}}
\end{figure}

The stable \ion{C}{3} profiles are not likely solely the result of problems with
the normalization of the spectra.  We confirm this 
by comparing the profiles of two nearby ISM lines in the 
three \fuse\ epochs.  
In Figure~\ref{fig:ism} we show two prominent but unsaturated 
ISM lines, \ion{S}{3}~\lam1012 and \ion{Ar}{1}~\lam1048,
in all three \fuse\ observation epochs.
We do see some normalization problems, particularly in the  \ion{Ar}{1}
region, likely attributable mainly to the low S/N in the 2002 data. 
The absorption depth in the 2002 \ion{Ar}{1}
profile is higher than in the other epochs when it is not expected to vary at all.
This may be a contributor to the apparent lack of variability in the 
\ion{C}{3} profiles.
We note, however, that the deepest absorption in the \ion{C}{3} profile peaks at 
-540 \kms, -460 \kms, and at -385 \kms. As discussed by S04,
the low ionization absorption at these velocities may not be part of 
the Mrk~279 outflow but may in fact be attributable to either the host galaxy
of the AGN or to a companion galaxy, MCG +12-13-024.

Photoionization models of quasar line emission
indicate that intermediate line region lies at approximately 1 pc
from the central ionizing source, a factor of $\sim$10 further than
than the broad line region
(Brotherton et al.\ 1994).
However, the UV and X-ray observations of the Seyfert galaxy NGC~4151 place the D+E subcomponents of its
outflow absorption in an accretion disk wind which is the ILR and which lies
at only 0.1 pc from the continuum source (Kraemer et al.\ 2006,
Crenshaw \& Kraemer 2007, see also Corbin 1997).
For Mrk~279, we have found ILR covering fractions that are consistently smaller than
those of the BLR and continuum, and in some cases $C^{\rm ILR}=0$.  This
would place the absorbing gas within or interior to the ILR.  Assuming that
the width of the intermediate lines, $\sim$2900~\kms, is produced gravitationally,
a black hole mass of $10^{7.54}$ M$_{\sun}$ (Peterson et al.\ 2004) gives a radius
of $\sim$0.02 pc for the ILR, an estimate that is only slightly larger than
measured constraints on the size of the BLR, $\sim$0.01 pc (Maoz et al.\ 1990).
We note, however, that this estimate of the ILR size is influenced by the 
relatively broad ILR emission features, certainly as compared with NGC~4151.  

\section{Summary}
We find that some of the variability, or lack thereof, in the absorption
profiles of the various species present in the \fuse\ and \stis\ data for Mrk~279 can
be understood in terms of changes in the effective covering fraction when all emission
components are included in the absorption model.  
Accounting for individual covering fractions in the continuum, broad lines, and intermediate
velocity width lines, we find less dramatic changes in the 
column densities of the highly ionized species are required by the absorption data than those 
that are predicted by photoionization modeling alone.  In particular, the contribution of
the covering fraction of the intermediate velocity width emission lines to the overall
effective covering fraction for a given ion and the changes in the relative contribution
of this component to the overall emission between the two epochs can account for a significant portion
of the changes observed in the absorption profiles.

If the \ion{C}{3} absorption is intrinsic to the outflow, the UV continuum flux variation between
2002 and 2003 is not commensurate with an absorber of very high ionization
parameter or of very large density, placing this absorbing structure at a substantial
distance from the ionizing source.  We consider this unlikely, however, given the
nonunity broad and intermediate emission line covering fractions derived for the intrinsic
absorption.

\acknowledgements
J.\ E.\ S.\ acknowledges the support of a National Research Council Associateship held at 
NASA Goddard Space Flight Center and the support of the Jess and Mildred Fisher
Endowed Chair in the Biological and Physical Sciences, held at Towson University.

\begin{deluxetable}{llccl}
\tablecolumns{4}
\tablewidth{22pc}
\tablecaption{Observations of MRK 279 \label{table-data}}
\tablehead{
\colhead{Instrument} &\colhead{ID} &\colhead{Start Date} &\colhead{Exp. (s)} }
\startdata
{\it FUSE} &P1080303     &1999-12-28         &61139   \\
{\it FUSE} &C0900201     &2002-05-18         &47414  \\
{\it FUSE} &D1540101     &2003-05-12         &91040  \\
STIS/E140M &O6JM01       &2002-05-18            &13193   \\
STIS/E140M &O8K10        &2003-05-13            &41386   \\
\enddata
\end{deluxetable}

\begin{deluxetable}{llccc}
\tablecolumns{5}
\tablewidth{29pc}
\tablecaption{Emission Line Fits to 2003 {\it FUSE} and STIS Spectra of Mrk~279
\label{table-emspec}}
\tablehead{
\colhead{Line} &\colhead{$\lambda_{\rm vac}$}
&\colhead{Flux\tablenotemark{1}}
&\colhead{Velocity\tablenotemark{2}}
&\colhead{FWHM} \\
\colhead{} &\colhead{(\AA)}
&\colhead{}
&\colhead{(\kms)}
&\colhead{(\kms)} }
\startdata
\ion{S}{6} &933.378 &4.09$\pm$1.64 &683$\pm$1782 &7569$\pm$192\\
\ion{S}{6} &944.523 &2.04$\pm$0.82 &683$\pm$1782 &7569$\pm$192\\
\ion{C}{3}  &977.020 &33.55$\pm$2.82 &384$\pm$458 &9206$\pm$146\\
\ion{N}{3} &989.799 &8.06$\pm$2.52 &417$\pm$1066 &9206$\pm$146\\
Ly$\beta$ broad &1025.7223 &31.52$\pm$2.47 &-79$\pm$7 &8731$\pm$61\\
Ly$\beta$ int. &1025.7223 &5.77$\pm$0.67 &-469$\pm$11 &3208$\pm$22\\
Ly$\beta$ narrow\tablenotemark{3} &1025.7223 &0.56$\pm$0.33 &145$\pm$22 &697\\
\ion{O}{6} broad &1031.9265 &71.66$\pm$2.58 &475$\pm$25 &7569$\pm$192\\
\ion{O}{6} int.  &1031.9265 &10.29$\pm$1.25 &-193$\pm$100 &2783$\pm$97\\
\ion{O}{6} broad &1037.6155 &35.83$\pm$1.29 &475$\pm$25 &7569$\pm$192\\
\ion{O}{6} int.  &1037.6155 &5.14$\pm$0.62 &-193$\pm$100 &2783$\pm$97\\
\ion{S}{4} int. &1062.66 &11.12$\pm$0.31 &-416$\pm$132 &3274$\pm$560\\
\ion{S}{4} int. &1072.97 &11.12$\pm$0.31 &-416$\pm$132 &3274$\pm$560\\
\ion{He}{2} &1085.15 &28.52$\pm$0.79 &-213$\pm$101 &7569$\pm$192\\
\ion{Si}{2} &1192.3  &3.57$\pm$0.79 &446$\pm$200 &9385$\pm$269\\
Ly$\alpha$ broad &1215.6701 &307.69$\pm$2.38 &-79$\pm$7 &8731$\pm$61\\
Ly$\alpha$ int.  &1215.6701 &120.23$\pm$0.52 &-469$\pm$11 &3208$\pm$22\\
Ly$\alpha$ narrrow\tablenotemark{3} &1215.6701 &7.09$\pm$0.48 &145$\pm$22 &697\\
\ion{N}{5} broad &1240.40 &106.66$\pm$2.80 &-56$\pm$53 &9908$\pm$19\\
\ion{N}{5} int.  &1240.40 &5.84$\pm$0.89 &-15$\pm$215 &2956$\pm$72\\
\ion{Si}{2} &1260.42 &5.32$\pm$0.61 &446$\pm$200 &9385$\pm$269\\
\ion{Si}{2}+\ion{O}{1} &1304.35 &9.82$\pm$0.89 &-69$\pm$140 &4020$\pm$732\\
\ion{C}{2} &1335.30  &1.68$\pm$0.59 &-219$\pm$352 &2422$\pm$207\\
\ion{Si}{4} &1393.76 &24.77$\pm$0.30 &-424$\pm$6 &9908$\pm$19\\
\ion{Si}{4} &1402.77 &12.38$\pm$0.15 &-424$\pm$6 &9908$\pm$19\\
\ion{O}{4}] &1402.06 &21.93$\pm$0.17 &-600$\pm$67 &3774$\pm$47\\
\ion{C}{4} broad &1549.05 &30.21$\pm$0.79 &4$\pm$19 &9908$\pm$19\\
\ion{C}{4} int.  &1549.05 &8.87$\pm$1.21 &-304$\pm$12 &2956$\pm$72\\
\ion{He}{2}  &1640.5 &80.96$\pm$3.30 &-11$\pm$122 &9909$\pm$19\\
\enddata
\tablenotetext{1}{Flux in units of 10$^{-14}$ ergs cm$^{-2}$ s$^{-1}$}
\tablenotetext{2}{Velocity relative to systemic redshift, $z=0.0305$}
\tablenotetext{3}{FWHM fixed to 2002 May value}
\end{deluxetable}

\clearpage

\clearpage

\clearpage

\clearpage

\clearpage

\clearpage

\clearpage

\clearpage

\clearpage

\clearpage
\begin{figure}
\epsscale{0.8}
\plotone{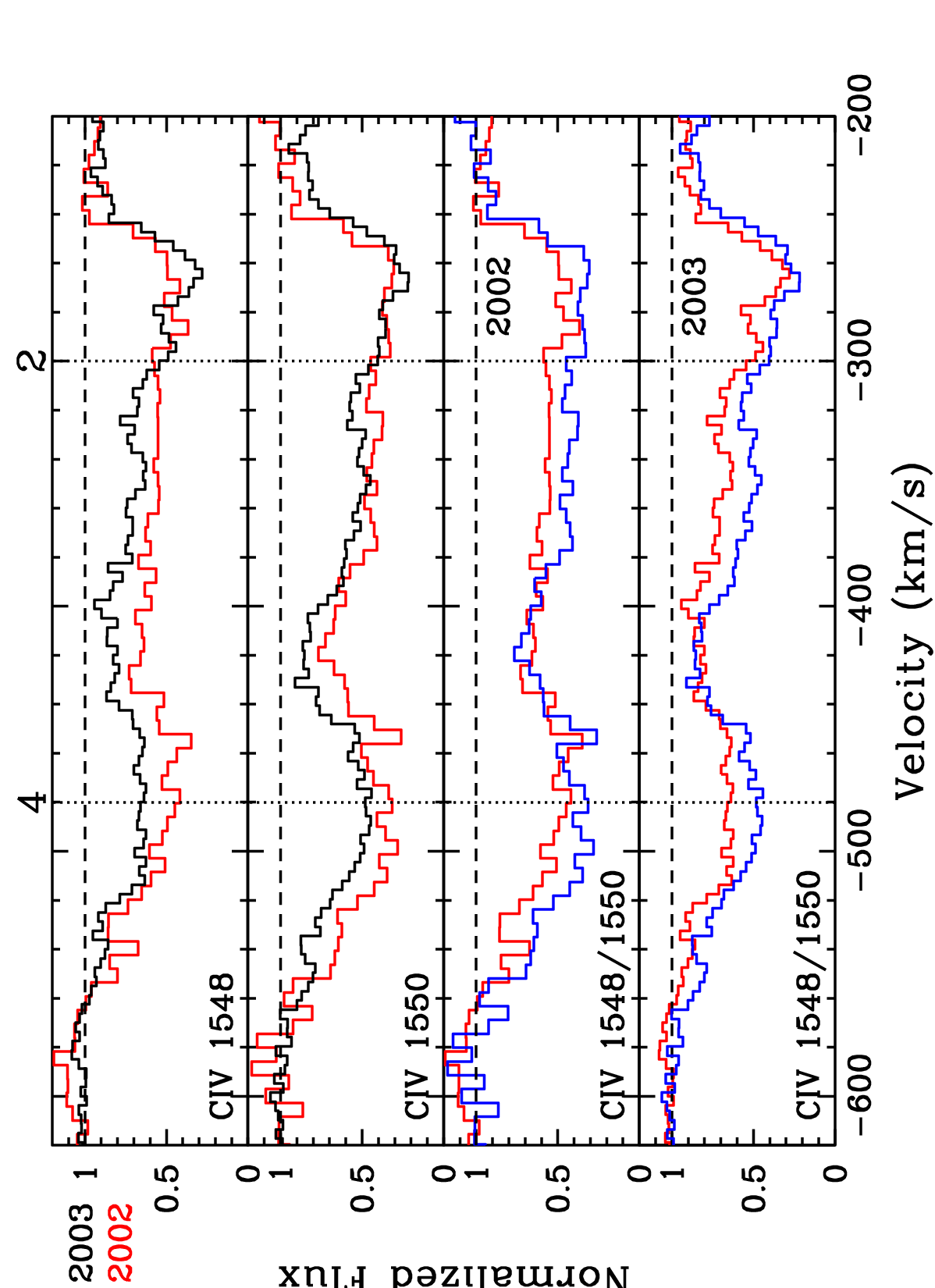}
\caption{Normalized \ion{C}{4} doublet over all observation epochs for
$C_{\rm ILR}= C_{\rm BLR}=C_{\rm cont}=1$. Positions of Components 2 and 4 are
labeled at top. Top panel shows the blue doublet components
in both the 2002 and 2003 epochs together, while the second panel shows this
for the red components.  Bottom three panels show the blue and red doublet components
together in each individual observation epoch.
\label{fig:c4doub}}
\end{figure}

\clearpage

\clearpage

\clearpage

\clearpage

\clearpage

\clearpage

\clearpage

\clearpage

\end{document}